# Nanometer-precision non-local deformation reconstruction using nanodiamond orientation sensing

Kangwei Xia[1+], Chu-Feng Liu[1+], Weng-Hang Leong[1+], Man-Hin Kwok[1], Zhi-Yuan Yang[1], Xi Feng[1], Ren-Bao Liu[1,2*], Quan Li[1,2*]

[1]. Department of Physics, The Chinese University of Hong Kong, Shatin, New Territories, Hong Kong, China

[2]. Centre for Quantum Coherence, The Chinese University of Hong Kong, Shatin, New Territories, Hong Kong, China.

[+]These authors contributed equally: Kangwei Xia, Chu-Feng Liu, Weng-Hang Leong.

[*]Correspondence and requests for materials should be addressed to R.-B.L. (email: rbliu@cuhk.edu.hk) or to Q.L. (email: liquan@phy.cuhk.edu.hk)

## Abstract

Information of material deformation upon loading is critical to evaluate mechanical properties of materials in general, and key to understand fundamental mechano-stimuli induced response of live systems in particular. Conventionally, such information is obtained at macroscopic scale about averaged properties, which misses the important details of deformation at individual structures; or at nanoscale with localized properties accessed, which misses the information of deformation at locations away from the force-loading positions. Here we demonstrate that an integration of nanodiamond orientation sensing and AFM nanoindentation offers a complementary approach to the nonlocal reconstruction of material deformation. This approach features a 5 nm precision in the loading direction and a sub-hundred nanometer lateral resolution. Using this tool, we mapped out the deformation of a PDMS thin film in air and that of the gelatin microgel particles in water, with precision high enough to disclose the significance of the surface/interface effects in the material deformation. The non-local nanometer-precision sensing of deformation upon a local impact brings in new opportunities in studying mechanical response of complex material systems.

Deformation measurements of heterogeneous nanomaterials[1,2] (e.g. nanocomposites and multiphase polymers) and live systems[3–5] (e.g. microbes and cells) are critical in evaluating their mechanical properties, and, more importantly, provide a viable means to study the mechano-stimulation response of live systems[6,7]. Nanoindentation[8], a method that gives the loading depth profile of materials, is commonly used to investigate the mechanical response upon loading[9,10]. With the development of pico-indenters and the advances of atomic force microscopy (AFM) based methods, localized deformation at the contact region can be routinely obtained and measured with lateral spatial resolution down to tens of nanometers [11–13]. However, these methods miss the information about the non-local response of the material to the local loading; such information is essential to understanding the deformation and impact transfer in heterogeneous structures, and how live systems respond to mechano-stimuli. Up to date, the only deformation measurement method that can extract non-local information is based on optical imaging[14–19]. The precision and the resolution of the optical measurement, however, is limited by the optical wavelength; or even with the super-resolution techniques based on single-molecule fluorescence, they are still limited to tens of nanometers. In typical optical measurement of indentation-induced deformation, indenters of large tip size (a few micrometers) are usually used to generate deformation large enough to be measurable (> few micrometers)[16–19]. Deformation of such large scales may miss the deformation characteristics at finer scale, which are essential to understanding many important deformation phenomena (such as the effects of surface tension on deformation at the interfaces between soft materials and liquids). Therefore, it is highly desirable to develop a new method that can have both non-local access of deformation and nanometer precision. Such a tool would facilitate the study of fine deformation characteristics of soft materials under weak external perturbation.

A new opportunity is diamond sensing[20–22]. Diamond sensing uses nitrogen vacancy (NV) centre[23] spins as the sensor, which can work under ambient conditions[24–28]. NV centre spin resonances are sensitive to small changes of magnetic field, making them effective quantum sensors to many physical parameters via direct or indirect magnetometry measurements[29–32]. In particular, since the spin resonance frequencies are shifted mostly by the magnetic field component projected to the nitrogen-vacancy axis, vector magnetometry has been developed to determine the relative orientation between the magnetic field and the crystallographic direction of the diamond lattice[26,28,33–35]. When a nanodiamond is docked on a material surface, the deformation is related to the orientation of the surface and hence the rotation of the nanodiamond. Based on this, with the combination of AFM-induced indentation, we propose and demonstrate a scheme of non-local deformation sensing using the rotation measurement of nanodiamonds docked on a surface.

We reconstructed the non-local surface topography using the three-dimensional rotation data of NDs docked on surfaces that were deformed by AFM indentation. We adopted two data-acquisition approaches – one with an ND fixed on a position and the AFM indentation position scanned around the ND (single-ND approach); another with a fixed indentation position and multiple NDs docked at various positions on the surface (multi-ND approach). In the proof-of-the-concept demonstrations, we employed the single -ND approach and reconstructed the deformation of a PDMS film with a 5 nm precision in the loading direction and an in-plane spatial resolution limited by the ND size (which is ~200 nm in our experiments). The measured mechanical property of the PDMS film is well described by a linear elastic model that has been established in literature[36], which validates the methodology. We further applied this sensing technique to measure deformation of gelatin microgel particles in water, using the multi-ND approach. The surface tension at soft materials/water interface is expected to play an important role in determining the deformation characteristics when the perturbation is weak[37].We measured for the first time the effect of surface tension at the soft material/water interface with nanoscale indentation. The load-responsive sensing protocols based on the ND rotation measurement provide a new approach to studying spatially resolved non-local deformation, which features a precision < 10 nm, bringing new opportunities to investigate mechanical properties of non-uniform media and potentially dynamical behaviors of systems under mechanical stimuli.

## Results

### Deformation reconstruction from ND rotation data

Figure 1a illustrates the measurement scheme. We consider the deformation of a material surface induced by a localized nanoindentation with an AFM tip. NDs are docked on the material surface. The deformation would cause the rotation of the NDs. The rotation of the NDs can be measured by optically detected magnetic resonance (ODMR) of the NV centres in the NDs under external magnetic fields.  The nonlocal deformation of the surface (at positions other than the indentation location) is thus reconstructed using the ND rotation data.

(1) Measurement of 3D rotation of NDs

The NV centres in an ND have four inequivalent crystallographic directions[23] ($\mathbf{NV}_{i,}\ i = 1,2,3,4$, along $(111)$, $(\bar{1}11)$, $(1\bar{1}1)$, and $(11\bar{1})$ correspondingly) as shown in Fig. 1b-1. The ground state of the NV centre is a spin triplet ($S = 1$) with a zero-field splitting $D \approx 2.87$ GHz between the $m_S = 0$ and $m_S = \pm 1$ states, which are quantized along the NV axis. In the presence of a weak magnetic field $\mathbf{B}$ ($< 100$ Gauss), the frequencies of the transitions $|0\rangle \leftrightarrow |\pm\rangle$ are $f_i^{\pm} \approx D \pm \gamma_e B \cos\psi_i$, where $\psi_i$ is the angle between the NV axis ($\mathbf{NV}_i$) and the magnetic field $\mathbf{B}$ (Fig. 1b-1), and $\gamma_e \approx 28$ MHz mT$^{-1}$ is the electron gyromagnetic ratio[21]. The angles $\psi_i$ and in turn the orientations of the NV axes are determined from the transition frequencies, which can be measured by the ODMR spectra. Note that the four angles $\psi_i$ of the NV centres and hence the ODMR frequencies would be identical if the ND is rotated about the magnetic field by any angle. To remove such ambiguity in determination of the orientation of the ND, we applied a second magnetic field $\mathbf{B}'$ so that the angles $\psi_i'$ between the NV axes and the magnetic field $\mathbf{B}'$ are determined from the new ODMR spectra (Fig. 1b-2). This way, the three-dimensional orientation of the ND is unambiguously determined. The rotation of the ND is characterized by the rotation axis $(\theta_{\mathrm{r}}, \phi_{\mathrm{r}})$, with $\theta_{\mathrm{r}}$ and $\phi_{\mathrm{r}}$ being the polar and azimuthal angles, and the rotation angle $\chi$ (Fig. 1b-3). The rotation axis and angle $(\theta_{\mathrm{r}}, \phi_{\mathrm{r}}, \chi)$ of the ND were deduced from the ODMR spectra before and after the rotation (details in Methods and SI Note 2). The rotation measurement protocol was first validated using a bulk diamond (with an ensemble of NV centres) of known crystallographic orientation and rotation operations (See SI Note 3).

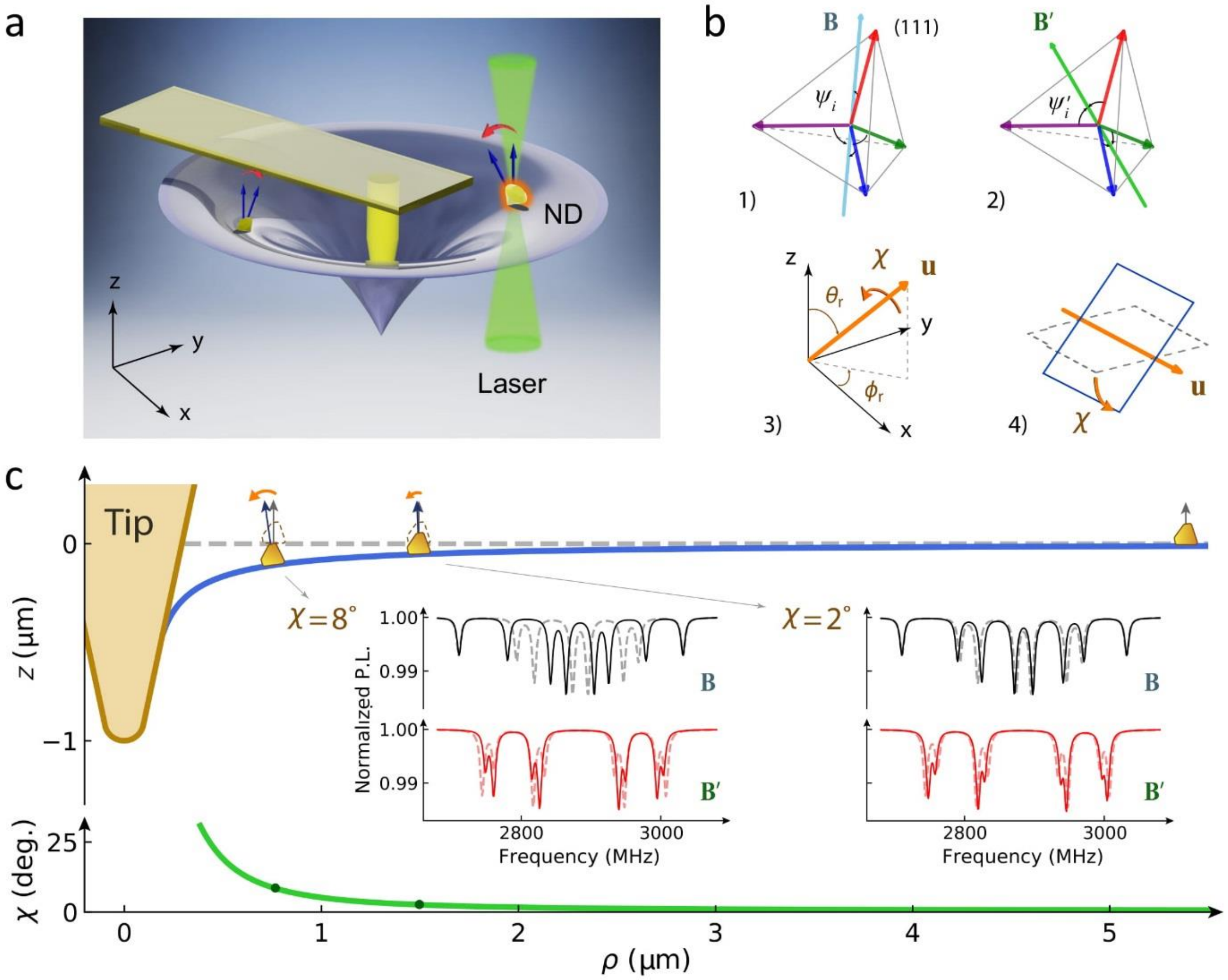

**Fig. 1 | Method of deformation reconstruction via measurement of the rotation of NDs docked on the surface. a,** An AFM tip imposes a localized indentation, causing non-local deformation of the surface. The NDs docked on the surface are rotated due to the deformation. **b, 1) & 2)** NV centres of four different orientations, with angles $\psi_i$ and $\psi_i'$ $(i = 1, 2, 3, 4)$ from two external magnetic fields **B** and **B′**, respectively. **3)** Rotation of the ND is characterized by the direction of the rotation axis $(\theta_r, \phi_r)$ and the rotation angle $\chi$. **4)** The tangential surfaces before (dotted) and after (solid) the deformation, in which the intersection and the angle between the two surfaces correspond respectively to the rotation axis and the rotation angle of the ND docked at the surface. **c,** Simulated deformation (blue line) upon an AFM tip indentation on a homogenous material. The green line shows the simulated rotation angles of NDs as functions of their distance from the AFM tip. Inset: Examples of the simulated OMDR spectra of the two NDs on the surface (indicated by arrows) under magnetic field **B** (black) or **B′** (red) with (solid line) or without (dash line) the indentation.

(2) Reconstruction of deformation

The deformation was reconstructed from the rotation of the NDs docked on the surface in the following procedure. The rotation of an ND was characterized by the rotation matrix $\mathbf{R}(\theta_{\mathrm{r}}, \phi_{\mathrm{r}}, \chi)$. The position of the ND is denoted as $\boldsymbol{\rho} = (\rho, \theta)$, written in polar coordinate. The relation between the rotation of the ND and the rotation of the tangential plane of the surface at $\boldsymbol{\rho}$, as illustrated in Fig. 1b-4, is given by the rotation of the normal direction, $\hat{\mathbf{n}}(\boldsymbol{\rho}) =$ $\mathbf{R}(\theta_r, \phi_r, \chi)\,\hat{\mathbf{z}}$, where $\hat{\mathbf{z}}$ is the normal direction of the surface before the indentation (the $z$-axis). The displacement of the surface at the position of the ND, denoted as $z(\boldsymbol{\rho})$, was reconstructed by the least-square fitting of the rotation data using the gradient field equation $\mathbf{g}(\boldsymbol{\rho}) = \nabla_{\boldsymbol{\rho}} z(\boldsymbol{\rho})$[38], where the gradient is related to the rotated normal direction by $\mathbf{n} = (-g_x, -g_y, 1)$ (for details see Methods). In reconstructing the deformation, we assumed that the attached NDs had no relative motion or rotation against the surface. This assumption was confirmed by the fact that the NDs recovered their original orientation after the indentation was released (see SI Fig. S11).

(3) ODMR simulation

To check the feasibility of the method, we ran numerical simulations of the ODMR spectra of NDs docked on the surface of a soft material under an AFM indentation (see Fig. 1c). We considered a homogenous material upon an AFM tip indentation (Fig. 1c) and adopted the linear elastic model (the Hertzian model)[39,40]. The parameters used in the simulations were: Young's modulus $E = 30$ kPa, and the load force $P = 10$ nN. The deformed surface and the rotation angle of the normal direction ($\chi$) are shown in blue and green lines in Fig. 1c, respectively. Taking two NDs at different distances from the AFM tip as examples, we simulated their ODMR spectra under **B** or **B′** with or without the indentation (see insets in Fig. 1c). In both NDs, eight ODMR resonance dips were found, corresponding to the +/- transition frequencies of the NV centres along the four crystallographic orientations. The frequency shifts were clearly seen between the spectra with (solid line) or without (dash line) the indentation, from which the rotation of the respective NDs was deduced.

**Non-local deformation reconstruction –single-ND method**

For a laterally homogeneous material, the deformation upon a nanoindentation can be reconstructed with an ND sensor fixed on a position and an AFM tip scanning around the ND. This single-ND method is possible because on the homogeneous material the deformation at position A upon an indentation at position B should be the same as the deformation at B upon an indentation at A.

We first demonstrated the single-ND method using a PDMS film. NDs with a typical size of $\sim 200$ nm were dispersed on the surface of PDMS film by spin-coating. The thickness of the PDMS film was $\sim 50$ μm with an oxidized surface layer formed by $O_2$ plasma treatment (for the preparation and the characterization of the PDMS film see Methods and SI Note 4). Figure 2a shows an AFM image of the PDMS surface with a single ND located in the centre $(0,0)$. The spatial correlation was built by comparing the AFM image and the confocal image taken from the same area, which helped to determine the coordinates of the ND in AFM topography image for further indentation experiments (details of the AFM-confocal correlation microscopy are given in SI Fig. S6).

The indentation experiment was carried out by descending the AFM tip into the PDMS film in the proximity of the ND with a programmed indentation sequence from location #1 to #92 (red dots in Fig. 2a). A constant loading force $P = 150$ nN was applied during indentation. The ODMR spectra of the ND under two different magnetic fields were measured with and without the indentation on each location from #1 to #92 (typical ODMR spectra plotted in SI Fig.

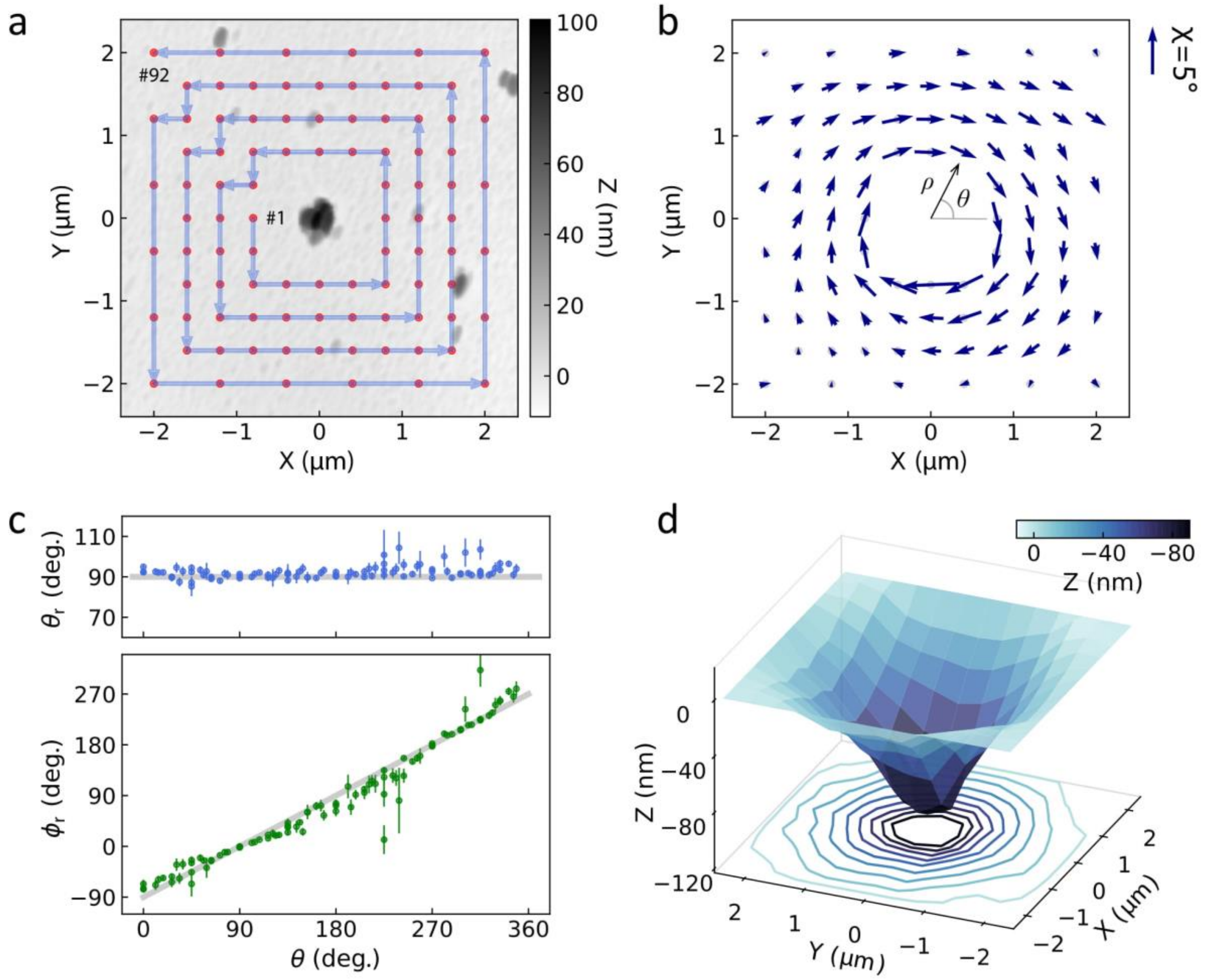

**Fig. 2 | Deformation reconstruction using the single-ND method. a,** AFM image of typical PDMS surface with an ND located at the centre (the origin). The red dots represent the indentation locations of the AFM tip, starting with #1 and ending with #92. **b**, The rotations of the ND plot as arrows for the displacements of the ND from the 92 indentation positions. The size of the arrows represents the magnitude of the rotation angles $\chi$ (scale bar shown on the right), and the direction is the rotation axes projected to the *x*-*y* plane. **c,** The polar angle $\theta_r$ and the azimuthal angle $\phi_r$ of the rotation axes of the ND as functions of $\theta$ for the 92 indentation positions. The grey line in the upper graph is $\theta_r = 90°$ and in the lower graph is $\phi_r = \theta - 90°$. **d,** The reconstructed surface of the PDMS film upon an AFM tip indentation at the centre (the origin).

S7). The rotations of the ND were derived from the ODMR spectra for various displacements $\boldsymbol{\rho}$ $(\rho, \theta)$ of the ND from the AFM indentation positions, shown in Figure (Fig. 2b), in which the arrows represent the projected rotation axes and the rotation angle. For all indentation locations, the polar angle $\theta_{\mathrm{r}}$ of the rotation axis was $\sim 90°$ and the azimuthal angle $\phi_{\mathrm{r}}$ was about $\phi_{\mathrm{r}} \sim \theta - 90°$ dependence (Fig. 2c), that is, the rotation axes of the ND were approximately on the *x*-*y* plane and perpendicular to the relative displacement $\boldsymbol{\rho}$ between the ND and the AFM indentation position.

The deformation due to the indentation by an AFM tip applied to the origin (0,0) was reconstructed from the rotation data of the ND as a function of the displacement from the AFM tip, as show in Fig. 2d, using the fact that the PDMS film is homogeneous on the *x*-*y* plane (see SI Fig. S5b). Figure 2d shows an axisymmetric characteristic about the origin.

We further show the rotation angle $\chi$ (Fig. 3a) and the reconstructed surface $z$ (Fig. 3b) as functions of the distance $\rho$ between the ND and the AFM tip. Similar trends were obtained in all measurements on NDs at different locations on PDMS (data of three NDs shown in Fig. 3 and SI Note 4), which suggests the homogeneity of the PDMS film.  The homogeneity assumption was also confirmed by the almost identical load-depth profiles obtained at different locations on the sample by AFM indentation (See SI Fig. S5b). The standard deviation of $\chi$ was $\sigma_\chi = 0.5°$ and in turn the standard deviation of the reconstructed *z* value was $\sigma_z = 5.2$ nm, which defines an unprecedented precision of deformation measurement (details of data fluctuation is given in SI Fig. S10).

The rotation and deformation data are well consistent with the numerical simulation based on a linear elastic bi-layer model[36]. The model consists of an oxidized surface layer of thickness $500$ nm on the top of bulk PDMS. The oxidation layer resulted from the plasma treatment of the PDMS sample[41,42]. The $500$ nm thickness of the oxidation layer was estimated from the AFM stiffness mapping of a cross-sectional PDMS sample (see SI Fig. S5a). In the simulation, the Young's modulus of the bulk PDMS was $\approx 0.78$ MPa (as measured by AFM indentation, see SI

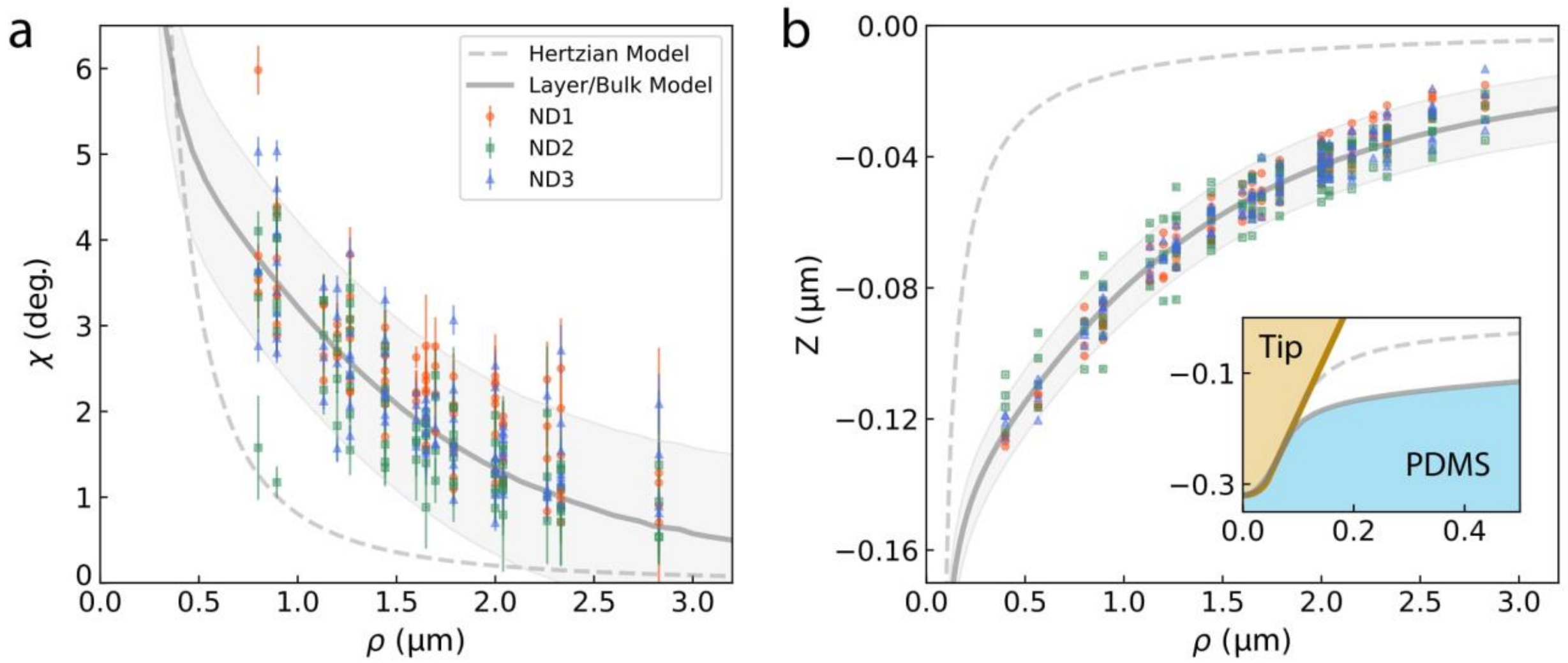

**Fig. 3 | Comparison between deformation reconstruction from experimental data and modelling.** **a,** The rotation angle ($\chi$) and **b,** the reconstructed deformation ($z$) as functions of the distance between the ND and the AFM tip. The solid lines are the simulation results of the bilayer model, with the shadowed regions bounded by $\chi \pm 2\sigma_\chi$ (1°) and $z \pm 2\sigma_z$ (10 nm) in **a** and **b**, respectively. The dashed lines are simulation results from the Hertzian model. The inset in **b** zooms in the simulation results near the AFM tip.

Note 4), and the Young's modulus of the oxidation layer was about 9.4 MPa (about 12 times greater than the bulk value, which is consistent with the literature[42]). The simulated rotation angle $\chi$ of the ND and the deformation $z$, plot respectively in Fig. 3a and Fig. 3b, are well consistent with the experimental data. In addition, the simulated depth-loading curve with the same parameters also matches the experimental result (see SI Fig. S12b). In contrast, the conventional Hertzian model, which does not contain the oxidation surface layer, led to simulation results (dash grey lines in Fig. 3) deviating significantly from the experimental data.

## Non-local deformation mapping – multi-ND method

The second protocol utilized multiple NDs to map the deformation around the indentation imposed at one spot. We demonstrated this multi-ND sensing of deformation induced by a single nanoindentation on a gelatin particle in water (Fig. 4a). By measuring the rotations of the NDs, the deformation of the gelatin particle was reconstructed.

We synthesized gelatin microparticles and dispersed the NDs on the surface in water (details see Methods). Figures 4b and 4c show the AFM and fluorescence images of a gelatin microparticle with NDs docked on the surface. The white dashed circle marks the boundary of the gelatin particle, which was about 30 μm in diameter and about 10 μm in height. The correlation between the AFM and the fluorescence images were used to identify the displacements of the NDs from an AFM indentation (details in SI Note 1). Figure 4d shows the fluorescence image of an examined sample area. The rotations of the NDs are plotted on the same image, with the directions of the arrows representing the rotation axes projected to the *x*-*y* plane and their lengths representing the rotation angles.

The rotation measurements were carried out on eight groups of NDs (Group I-VIII) around eight different indentation locations. In all cases, the rotation axes of all measured NDs were found approximately on the *x*-*y* plane and the NDs all rotated towards their respective indentation locations (see SI Fig. S14). Figures 4e and 4f plot the rotation angles of the NDs in

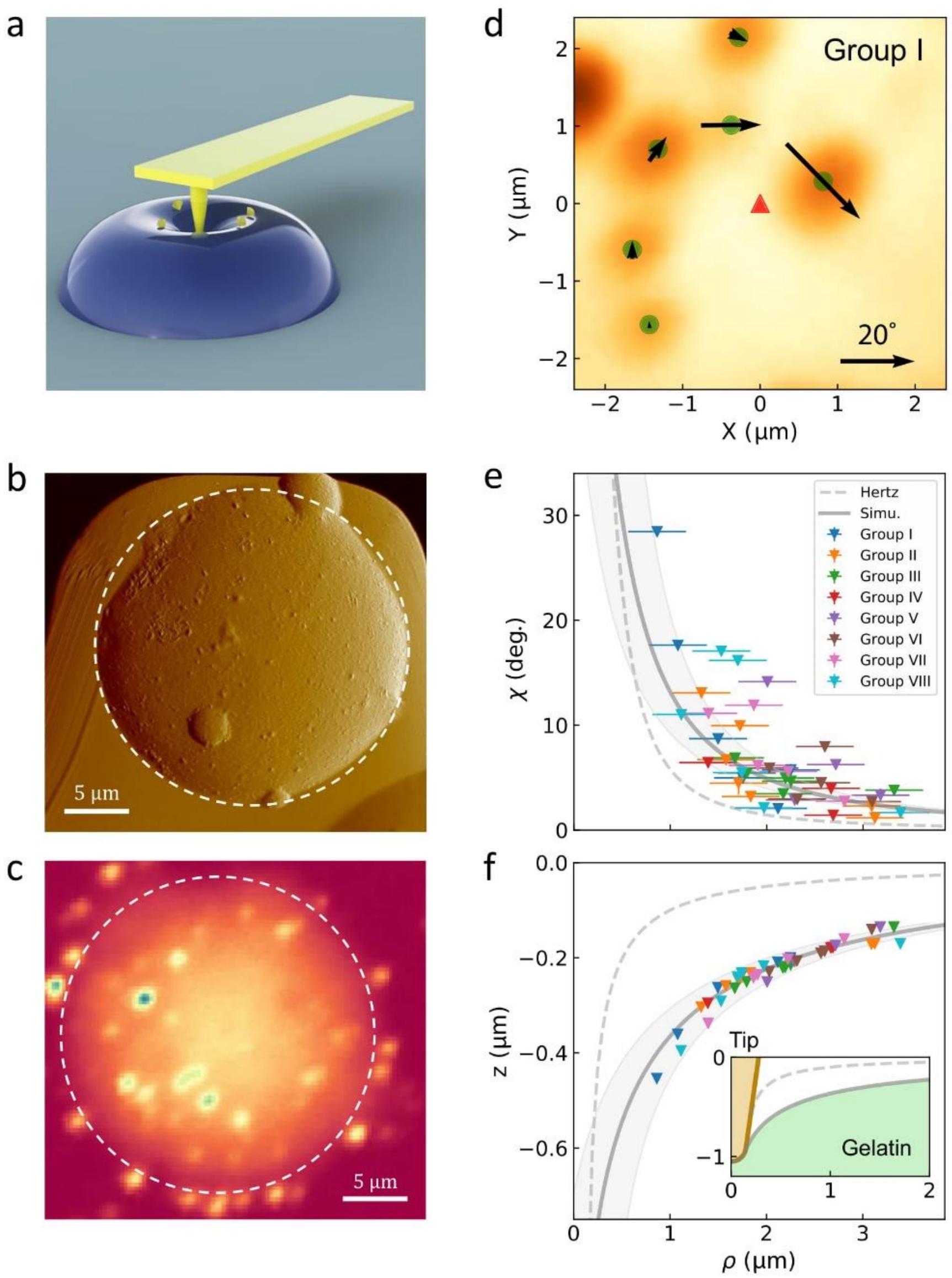

**Fig. 4 | Multi-ND mapping of deformation induced by a single indentation. a,** Schematic of the gelatin deformation by an AFM indentation. The NDs were dispersed on the surface of the gelatin. **b,** The AFM image, and **c,** The fluorescence confocal image of the gelatin sample. **d,** Fluorescence confocal image showing the NDs in Group I. The green dots mark the NDs indicated by the highest fluorescence intensity. The directions and the sizes of the arrows represent the rotation axes and the rotation angles of the NDs, respectively. The red triangle marks the position of the AFM tip indentation. **e,** The rotation angle $\chi$ and **f,** the reconstructed deformation taken from all eight groups of NDs (around eight different AFM indentation positions), plotted as functions of the distance between the NDs and the AFM tip indentation positions. The solid grey lines are the simulation results using the surface tension model (with a $\pm 300$ nm grey shadow given by the optical resolution). The dashed lines are simulation results from the Hertzian model. The inset zooms in the simulation results of the gelatin surface near the AFM tip.

Group I-VIII and the corresponding deformation upon the respective indentations, as functions of the distance between the NDs and the indentation locations (see Methods). The data from the NDs in the same group are presented in the same color.

The high precision and the small AFM tip allowed us to study an important effect in deformation upon loading – the effect of surface tension. The surface tension has recently been identified as a key factor in affecting material deformation upon loading[16–19,37]. The elastocapillary phenomenon has been disclosed as an important reflection of competition between the bulk elastic energy and surface energy[37]. However, the elastocapillary phenomenon only becomes significant when the scale of deformation is smaller than the elastocapillary length scale $L = \tau^0/E$ ($\tau^0$ is the surface tension and $E$ is the Young's modulus). To have an observable elastocapillary effect, one either requires the materials be very soft (with low $E$) and/or have high surface tension (large $\tau^0$), or is able to measure a small deformation to high precision (for small indentation caused by a small tip with a weak loading).

In our experiment, the indentation was induced by an AFM tip with a tip radius $\sim 0.1$ μm and with a small loading force of 15 nN. The deformation measurement based on ND rotation sensing allowed the small deformation to be measured to a high precision. Thus, we were able to perform a quantitative comparison between the experimental data and the elastocapillary model. We employed an elastic model including the effect of the surface tension to simulate the gelatin deformation upon an AFM indentation[43] (see Methods and SI Note 7). The simulated results are shown in Figs. 4e and 4f. The simulation assumed the Young's modulus $E = 6.5$ kPa and a surface tension $\tau^0 = 6.5$ mN m$^{-1}$ for the gelatin particle measured in water. These parameters were consistent with the literature reports of similar material systems[44,45] and were also confirmed by agreement between the experimental data and the numerical simulation of the depth-loading profile (Fig. S17b). The surface tension effect was clearly observed in our deformation measurements. Indeed, the simulation without the surface tension effect (Hertzian model, grey dashed lines in Figs. 4e and 4f) deviates significantly from the experimental data.

## Discussion and conclusion

In conclusion, we have developed a sensitive method to measure nonlocal nanoscale deformation by combining nanodiamond orientation sensing and AFM indentation. The developed technique features a high precision ($\sim 5$ nm) in deformation along the loading direction, which is one or two orders of magnitude higher than that of the conventional fluorescence-based methods. The deformed surface profile obtained with high precision contains abundant information of material mechanical properties, leading to the disclosure of heterostructures in PDMS films (due to surface oxidation), and the first measurement of the elastocapillary effects at the gelatin/water interface.

The two protocols, namely, the single-ND method and the multi-ND method, have complementary advantages and disadvantages. The single-ND approach provides an unprecedentedly high precision ($\sim 5$ nm) in the direction of loading ($z$) with a high laterally spatial resolution ($\sim 200$ nm). However, it relies on the assumption that the material is laterally homogeneous. For applications on laterally inhomogeneous materials, the multi-ND method is needed. The multi-ND method suffers from limited in-plane spatial resolution (depending on how dense NDs can be dispersed on the surface for ODMR measurement).

The methods developed in this paper are expected to offer new approaches to investigating surface effect of the materials as well as live systems in liquid environments upon external weak perturbation.

## Methods

### Setup and sample preparation.

We fabricated an omega shape microwave antenna on a glass slide and then spin coated PDMS on the top. The thickness of the PDMS was $\approx 50$ μm. When the PDMS was de-gassed, we modified the PDMS surface with $O_2$ plasma. The dilute ND solution (2 μg $mL^{-1}$ concentration) was thus spin coated on the PDMS film.

Gelatin (1.5 g) was dissolved in $10$ mL deionized water at $60$ °C. The $0.15$ g $mL^{-1}$ gelatin solution was added to the Span80 solution and the mixture was vigorously shaken for 20 s. After the shaking, the gelatin in toluene emulsion was stirred at 300 rpm and cooled under ambient conditions for 1 h. Then, the solidified particles were sonicated for 2 min. Finally, the particles were washed three times with 30 mL acetone, and then three times with 30 mL deionized water. The synthesized Gelatin microspheres were spin coated onto a confocal dish with a microwave antenna ($20$ μm copper wire was bonded on the surface of a glass slide before a holder was glued on the glass slide to have a liquid chamber with microwave access). Then the ND solution was dispersed on gelatin microspheres using the drop casting method.

The setup was a home-built confocal microscope correlated with an atomic force microscope (see SI Note 1 for details). In the PDMS indentation experiments, two magnetic fields were formed by combinations of two permanent magnets (one was fixed, and the other one was movable between two positions so that the magnetic field was switchable between $\mathbf{B}$ and $\mathbf{B}'$). In the gelatin indentation experiments, the magnetic field switching was implemented by a permanent magnet and an electric magnet (in which the current alternated its direction and kept its magnitude constant to avoid heating difference).

## Tracking ND rotation by ODMR spectra

The normalized ODMR spectra under the two known magnetic fields are written as[21]

$$S(f) = b - \sum_i C_i \left[\frac{\Delta f^2}{4\left(f - f_i^+\right)^2 + \Delta f^2} + \frac{\Delta f^2}{4\left(f - f_i^-\right)^2 + \Delta f^2}\right], \quad (1)$$

where $b$ is the baseline, $C_i$ represents the ODMR contrast of the NV centres along the $i$th crystallographic orientation, $\Delta f$ is the linewidth (FWHM), and $f_i^{\pm}$ are the transition frequencies. Instead of fitting the transition frequencies, we firstly fit the ODMR spectra under the two magnetic fields before the rotation, with the initial ND orientation (denoted as the Euler angles $(\alpha, \beta, \gamma)$, see SI Note 2-1 and Fig. S2b), the baselines, the contrasts, the FWHM, and the shift of the zero-field splitting $\Delta D$ (slightly different among different NDs) being the fitting parameters. Then, the rotation of the ND were determined by fitting the two ODMR spectra after the rotation with the specific rotation axis and angle $(\theta_{\mathrm{r}}, \phi_{\mathrm{r}}, \chi)$, the contrasts, and the baselines as fitting parameters. The Euler angles, the FWHM, and $\Delta D$ were fixed to be the results of the before rotation fitting. The fitting results of the contrasts and the baselines were approximately identical for the rotation measurements of the same ND. For details of the fitting methods, see SI Note 2-2.

**Deformation reconstruction algorithm**

The deformed surface $z(\boldsymbol{\rho})$ was reconstructed by the least-square fitting of the ND rotation data using the gradient field $\mathbf{g}(\boldsymbol{\rho}) = \boldsymbol{\nabla}_{\boldsymbol{\rho}} z(\boldsymbol{\rho})$[38]. The normal vector of the deformed surface is $\mathbf{n} = (-g_x, -g_y, 1)$ , which is related to the rotation of the attached ND as $\hat{\mathbf{n}}(\boldsymbol{\rho}) = \mathbf{R}(\theta_\mathrm{r}, \phi_\mathrm{r}, \chi)\, \hat{\mathbf{z}}$, where $\mathbf{R}$ is the rotation matrix with rotation axis $(\theta_\mathrm{r}, \phi_\mathrm{r})$ and rotation angle $\chi$, and $\hat{\mathbf{z}}$ is the $z$-axis. Therefore, the gradient field was deduced from the rotation of the ND by $\mathbf{g}(\boldsymbol{\rho}) = (-\frac{\hat{n}_x}{\hat{n}_z}, -\frac{\hat{n}_y}{\hat{n}_z})$. For the gelatin experiment, with the assumption that the deformation was axisymmetric, the measured gradient field in polar coordinate was $\left(g_\rho, g_\theta\right) = (\tan \chi(\rho), 0)$. The deformed surface was reconstructed by integration $z(\rho) = \int_0^\rho \tan \chi(\rho')\, d\rho'$. For details, see SI Note 4 and 6.

**Numerical simulation of deformation**.

The indented deformation of the PDMS film was simulated with a half-infinite linear elastic model with a harder coating layer under an axisymmetric loading[36]. The Young's modulus of the PDMS and the oxidation layer were chosen as 0.78 MPa and 9.4 MPa, respectively, the Poisson's ratio was 0.33, and the thickness of the oxidation layer was 500 nm. The chosen parameters are consistent with the literatures[42,46] and our AFM measurement, for details see SI Note 5.

In the gelatin case, we used a half-infinite axisymmetric linear elastic model including the effect of the surface tension[43]. The Young's modulus was chosen as 6.5 kPa, the Poisson's ratio was 0.5, and the surface tension of the water-gelatin interface was 6.5 mN $m^{-1}$. The chosen parameters agree with the literatures[44,45], for details see SI Note 7.

**Data availability.** The data which supports the findings of this work is available upon request from the corresponding authors.

## References

1. Tai, K., Dao, M., Suresh, S., Palazoglu, A. & Ortiz, C. Nanoscale heterogeneity promotes energy dissipation in bone. *Nat. Mater.* **6,** 454–462 (2007).

2. Nicolas, A., Ferrero, E. E., Martens, K. & Barrat, J.-L. Deformation and flow of amorphous solids: Insights from elastoplastic models. *Rev. Mod. Phys.* **90,** 045006 (2018).

3. Maître, J.-L., Salbreux, G., Jülicher, F., Paluch, E. & Heisenberg, C. Adhesion Functions in Cell Sorting by of Adhering Cells. *Science* **253,** 253–257 (2012).

4. Style, R. W. *et al.* Traction force microscopy in physics and biology. *Soft Matter* **10,** 4047–4055 (2014).

5. Rigato, A., Miyagi, A., Scheuring, S. & Rico, F. High-frequency microrheology reveals cytoskeleton dynamics in living cells. *Nat. Phys.* **13,** 771–775 (2017).

6. Desmaële, D., Boukallel, M. & Régnier, S. Actuation means for the mechanical stimulation of living cells via microelectromechanical systems: A critical review. *J. Biomech.* **44,** 1433–1446 (2011).

7. Septiadi, D., Crippa, F., Moore, T. L., Rothen-Rutishauser, B. & Petri-Fink, A. Nanoparticle-Cell Interaction: A Cell Mechanics Perspective. *Adv. Mater.* **30,** 1704463 (2018).

8. Fischer-Cripps, A. C. *Nanoindentation*. (Springer New York, 2004).

9. VanLandingham, M. R., Villarrubia, J. S., Guthrie, W. F. & Meyers, G. F. Nanoindentation of polymers: an overview. *Macromol. Symp.* **167,** 15–44 (2001).

10. Ebenstein, D. M. & Pruitt, L. A. Nanoindentation of biological materials. *Nano Today* **1,** 26–33 (2006).

11. Adamcik, J., Berquand, A. & Mezzenga, R. Single-step direct measurement of amyloid fibrils stiffness by peak force quantitative nanomechanical atomic force microscopy. *Appl. Phys. Lett.* **98,** 193701 (2011).

12. Passeri, D., Rossi, M., Tamburri, E. & Terranova, M. L. Mechanical characterization of polymeric thin films by atomic force microscopy based techniques. *Anal. Bioanal. Chem.* **405,** 1463–1478 (2013).

13. Dufrêne, Y. F., Martínez-Martín, D., Medalsy, I., Alsteens, D. & Müller, D. J. Multiparametric imaging of biological systems by force-distance curve–based AFM. *Nat. Methods* **10,** 847–854 (2013).

14. Jerison, E. R., Xu, Y., Wilen, L. A. & Dufresne, E. R. Deformation of an Elastic Substrate by a Three-Phase Contact Line. *Phys. Rev. Lett.* **106,** 186103 (2011).

15. Style, R. W. *et al.* Universal Deformation of Soft Substrates Near a Contact Line and the Direct Measurement of Solid Surface Stresses. *Phys. Rev. Lett.* **110,** 066103 (2013).

16. Style, R. W., Hyland, C., Boltyanskiy, R., Wettlaufer, J. S. & Dufresne, E. R. Surface tension and contact with soft elastic solids. *Nat. Commun.* **4,** 2728 (2013).

17. Jensen, K. E., Style, R. W., Xu, Q. & Dufresne, E. R. Strain-Dependent Solid Surface Stress and the Stiffness of Soft Contacts. *Phys. Rev. X* **7,** 041031 (2017).

18. Pham, J. T., Schellenberger, F., Kappl, M. & Butt, H.-J. From elasticity to capillarity in soft materials indentation. *Phys. Rev. Mater.* **1,** 015602 (2017).

19. Xu, Q. *et al.* Direct measurement of strain-dependent solid surface stress. *Nat. Commun.* **8,** 555 (2017).

20. Schirhagl, R., Chang, K., Loretz, M. & Degen, C. L. Nitrogen-Vacancy Centers in Diamond: Nanoscale Sensors for Physics and Biology. *Annu. Rev. Phys. Chem.* **65,** 83–105 (2014).

21. Rondin, L. *et al.* Magnetometry with nitrogen-vacancy defects in diamond. *Reports Prog. Phys.* **77,** 056503 (2014).

22. Casola, F., van der Sar, T. & Yacoby, A. Probing condensed matter physics with magnetometry based on nitrogen-vacancy centres in diamond. *Nat. Rev. Mater.* **3,** 17088 (2018).

23. Doherty, M. W. *et al.* The nitrogen-vacancy colour centre in diamond. *Phys. Rep.* **528,** 1–45 (2013).

24. Fu, C.-C. *et al.* Characterization and application of single fluorescent nanodiamonds as cellular biomarkers. *Proc. Natl. Acad. Sci.* **104,** 727–732 (2007).

25. Neugart, F. *et al.* Dynamics of diamond nanoparticles in solution and cells. *Nano Lett.* **7,** 3588–3591 (2007).

26. McGuinness, L. P. *et al.* Quantum measurement and orientation tracking of fluorescent nanodiamonds inside living cells. *Nat. Nanotechnol.* **6,** 358–363 (2011).

27. Kucsko, G. *et al.* Nanometre-scale thermometry in a living cell. *Nature* **500,** 54–58 (2013).

28. Le Sage, D. *et al.* Optical magnetic imaging of living cells. *Nature* **496,** 486–489 (2013).

29. Taylor, J. M. *et al.* High-sensitivity diamond magnetometer with nanoscale resolution. *Nat. Phys.* **4,** 810–816 (2008).

30. Cai, J., Jelezko, F. & Plenio, M. B. Hybrid sensors based on colour centres in diamond and piezoactive layers. *Nat. Commun.* **5,** 4065 (2014).

31. Wang, N. *et al.* Magnetic Criticality Enhanced Hybrid Nanodiamond Thermometer under Ambient Conditions. *Phys. Rev. X* **8,** 011042 (2018).

32. Zhang, T. *et al.* Hybrid nanodiamond quantum sensors enabled by volume phase transitions of hydrogels. *Nat. Commun.* **9,** 3188 (2018).

33. Steinert, S. *et al.* High sensitivity magnetic imaging using an array of spins in diamond. *Rev. Sci. Instrum.* **81,** 043705 (2010).

34. Doherty, M. W. *et al.* Measuring the defect structure orientation of a single NV centre in diamond. *New J. Phys.* **16,** 063067 (2014).

35. Dovzhenko, Y. *et al.* Magnetostatic twists in room-temperature skyrmions explored by nitrogen-vacancy center spin texture reconstruction. *Nat. Commun.* **9,** 2712 (2018).

36. Li, J. & Chou, T.-W. Elastic field of a thin-film/substrate system under an axisymmetric loading. *Int. J. Solids Struct.* **34,** 4463–4478 (1997).

37. Style, R. W., Jagota, A., Hui, C.-Y. & Dufresne, E. R. Elastocapillarity: Surface Tension and the Mechanics of Soft Solids. *Annu. Rev. Condens. Matter Phys.* **8,** 99–118 (2017).

38. Harker, M. & O'Leary, P. Regularized Reconstruction of a Surface from its Measured Gradient Field. *J. Math. Imaging Vis.* **51,** 46–70 (2015).

39. Hertz, H. Ueber die Berührung fester elastischer Körper. *J. für die reine und Angew. Math. (Crelle's Journal)* **1882,** 156–171 (1882).

40. Sneddon, I. N. The relation between load and penetration in the axisymmetric boussinesq problem for a punch of arbitrary profile. *Int. J. Eng. Sci.* **3,** 47–57 (1965).

41. Bowden, N., Huck, W. T. S., Paul, K. E. & Whitesides, G. M. The controlled formation of ordered, sinusoidal structures by plasma oxidation of an elastomeric polymer. *Appl. Phys. Lett.* **75,** 2557–2559 (1999).

42. Mills, K. L., Zhu, X., Takayama, S. & Thouless, M. D. The mechanical properties of a surface-modified layer on polydimethylsiloxane. *J. Mater. Res.* **23,** 37–48 (2008).

43. Long, J. M. & Wang, G. F. Effects of surface tension on axisymmetric Hertzian contact problem. *Mech. Mater.* **56,** 65–70 (2013).

44. Radmacher, M., Fritz, M. & Hansma, P. K. Imaging soft samples with the atomic force microscope: gelatin in water and propanol. *Biophys. J.* **69,** 264–270 (1995).

45. Bialopiotrowicz, T. & Jańczuk, B. Surface properties of gelatin films. *Langmuir* **18,** 9462–9468 (2002).

46. Armani, D., Liu, C. & Aluru, N. Re-configurable fluid circuits by PDMS elastomer micromachining. in *Technical Digest. IEEE International MEMS 99 Conference. Twelfth IEEE International Conference on Micro Electro Mechanical Systems (Cat. No.99CH36291)* 222–227 (IEEE, 1999).

## Acknowledgements

We would like to thank the discussion with Ting Zhang for the gelatin synthesis; and Sen Yang for the setup construction. This work was supported by Hong Kong RGC/CRF (C4006-17G), and CUHK Group Research Scheme under project code 3110126.

## Author contributions

Q.L. and R.B.L. conceived the idea and supervised the project. K. X, C.F.L, W. H. L, R. B. L. and Q. L. designed the experiments. K. X., C. F. L. constructed the setup. C. F. L. and K. X. performed the experiments. W. H. L., C. F. L., K. X., R. B. L. and Q. L. analyzed the data. W. H. L. designed the rotation tracking protocol and the reconstruction algorithm, and performed the numerical simulation. M. H. K. synthesized the gelatin. Z. Y. Y. and X. F. synthesized the PDMS film. K. X., C. F. L., W. H. L., R. B. L. and Q. L. wrote the paper and all authors commented on the manuscript.

**Competing interests.** The authors declare no competing interests.

**Materials & Correspondence.** Correspondence and requests for materials should be addressed to R.-B.L. (email: rbliu@cuhk.edu.hk) or to Q.L. (email: liquan@phy.cuhk.edu.hk)

# Supplementary Information

*for*

# Nanometer-precision non-local deformation reconstruction using nanodiamond orientation sensing

Kangwei Xia[1+], Chu-Feng Liu[1+], Weng-Hang Leong[1+], Man-Hin Kwok[1], Zhi-Yuan Yang[1], Xi Feng[1], Ren-Bao Liu[1,2*], Quan Li[1,2*]

[1]. Department of Physics, The Chinese University of Hong Kong, Shatin, New Territories, Hong Kong, China

[2]. Centre for Quantum Coherence, The Chinese University of Hong Kong, Shatin, New Territories, Hong Kong, China.

[+]These authors contributed equally: Kangwei Xia, Chu-Feng Liu, Weng-Hang Leong.

[*]Correspondence and requests for materials should be addressed to R.-B.L. (email: rbliu@cuhk.edu.hk) or to Q.L. (email: liquan@phy.cuhk.edu.hk)

## Contents

**Supplementary Information Note 1. Setup**

We reconstructed a confocal-atomic force microscopy (AFM) correlation microscope to enable the nonlocal deformation reconstruction via nanodiamond (ND) rotation measurement (Fig. S1). The optically detected magnetic resonance (ODMR) measurements were carried out using a home-built laser scanning confocal microscope. A 532 nm laser was adopted (MGL-III-532-200 mW, CNI), and reduction of its power fluctuation down to 0.1% was realized by applying a proportional integral derivative (PID) feedback control. A Nikon 100x (1.45 NA) oil immersion objective lens was used to collect the ND's fluorescence signals, which was then detected by an avalanche photodiode (APD, SPCM-AQRH-15-FC, Excelitas) and counted by NIDAQ (PCIe-6363, National Instrument).

A microwave (MW) source (N5171B EXG Signal Generator, Keysight) and an amplifier (ZHL-16W-43-S+, Mini-Circuit) were used to generate microwave for ODMR spectrum acquisition. To achieve high microwave efficiency, the MW was delivered to the NDs by an omega microstructure (diameter, 800 μm).

The AFM scanning head (BioScope Resolve, Bruker) was mounted on the confocal microscope to measure the topography and perform indentation. A shielding was built to isolate the confocal-AFM correlated microscope from external environment using sound-absorbing foams and a temperature PID control module was used to keep the temperature fluctuation $< 0.1$ ℃. In this way, sample drifting was minimized during the indentation experiment.

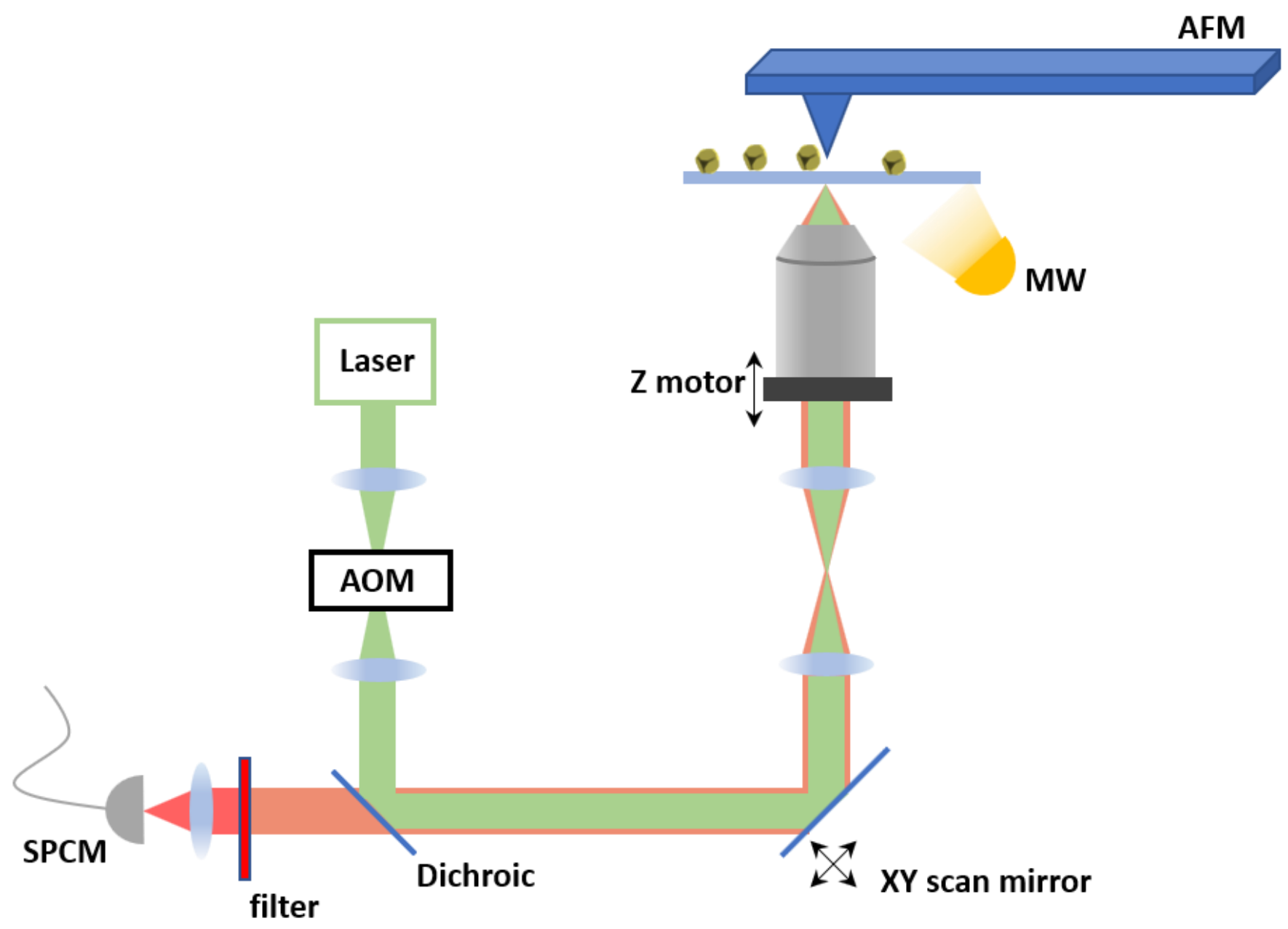

**Fig. S1 |** Schematic of the setup. AOM: acousto-optic modulator, MW: microwave, AFM: atomic force microscopy, SPCM: single photon counting modules.

A peak force tapping mode was applied to obtain the topography information of the samples. Before the indentation experiments, correlating the confocal system and the AFM system was realized in two steps. We first manually overlapped the excitation laser spot with the AFM tip under the top view camera on top of the AFM. Then, an AFM topography image and a fluorescence image were captured and a correlation between the two was established by comparing the patterns of the NDs distribution in the two images.

In the gelatin indentation experiments, the size (height) of gelatin particles (~10 μm) is much larger than the size of NDs (~ hundreds of nanometers), which made it difficult to resolve the NDs on the gelatin in the AFM image. The correlation between the AFM and the fluorescence images was established by overlapping the fluorescence image and AFM image of NDs on the substrate(cover slide). For the NDs on the gelatin particles, their coordinates were determined from the fluorescence images instead of the AFM ones. Consequently, the spatial resolution of the gelatin deformation measurements was limited by the optical resolution (~300 nm).

## Supplementary Information Note 2. Method of non-local deformation measurement

### (1) ND orientation and rotation

There are four possible NV axes in a diamond lattice, $(111)$, $(1\bar{1}\bar{1})$, $(\bar{1}1\bar{1})$ and $(\bar{1}\bar{1}1)$[1], as shown in Fig. S2a. The orientation of an ND can be unambiguously defined using the NV axes. We first use the orientation of a certain ND (labelled as ND0) to establish the laboratory coordinates (Fig. S2a) by choosing the $(111)$ direction ($\mathbf{NV}_1^0$) as the *z*-axis of and the $(1\bar{1}\bar{1})$ direction ($\mathbf{NV}_2^0$) in the *x*-*z* plane. The four NV orientations $\mathbf{NV}_i^0$ of ND0 can then be written as unit vectors as

$$\mathbf{NV}_1^0 = (0,\ 0,\ 1), \qquad \mathbf{NV}_2^0 = \frac{1}{3}\left(2\sqrt{3},\ 0,\ -1\right),$$

$$\mathbf{NV}_3^0 = \frac{1}{3}\left(-\sqrt{3},\ -\sqrt{6},\ -1\right), \qquad \mathbf{NV}_4^0 = \frac{1}{3}\left(-\sqrt{3},\ \sqrt{6},\ -1\right). \qquad \text{(S2-1)}$$

The orientation of an arbitrary ND (ND-I) is specified by the Euler angles $(\alpha, \beta, \gamma)$ (Fig. S2b), the angles of three sequential rotations that transfer the orientation of ND0 (whose crystallographic directions set the laboratory coordinate system) to that of ND-I: first a rotation of $\alpha$ about the $z$-axis in laboratory coordinate system (which transfers the $y$-axis to $\mathbf{N}(y)$); then a rotation of $\beta$ about $\mathbf{N}(y)$ (which transfers the $z$-axis to the $z'$-axis); and finally a rotation of $\gamma$ about the $z'$-axis. The rotation matrix of such an operation is[2]

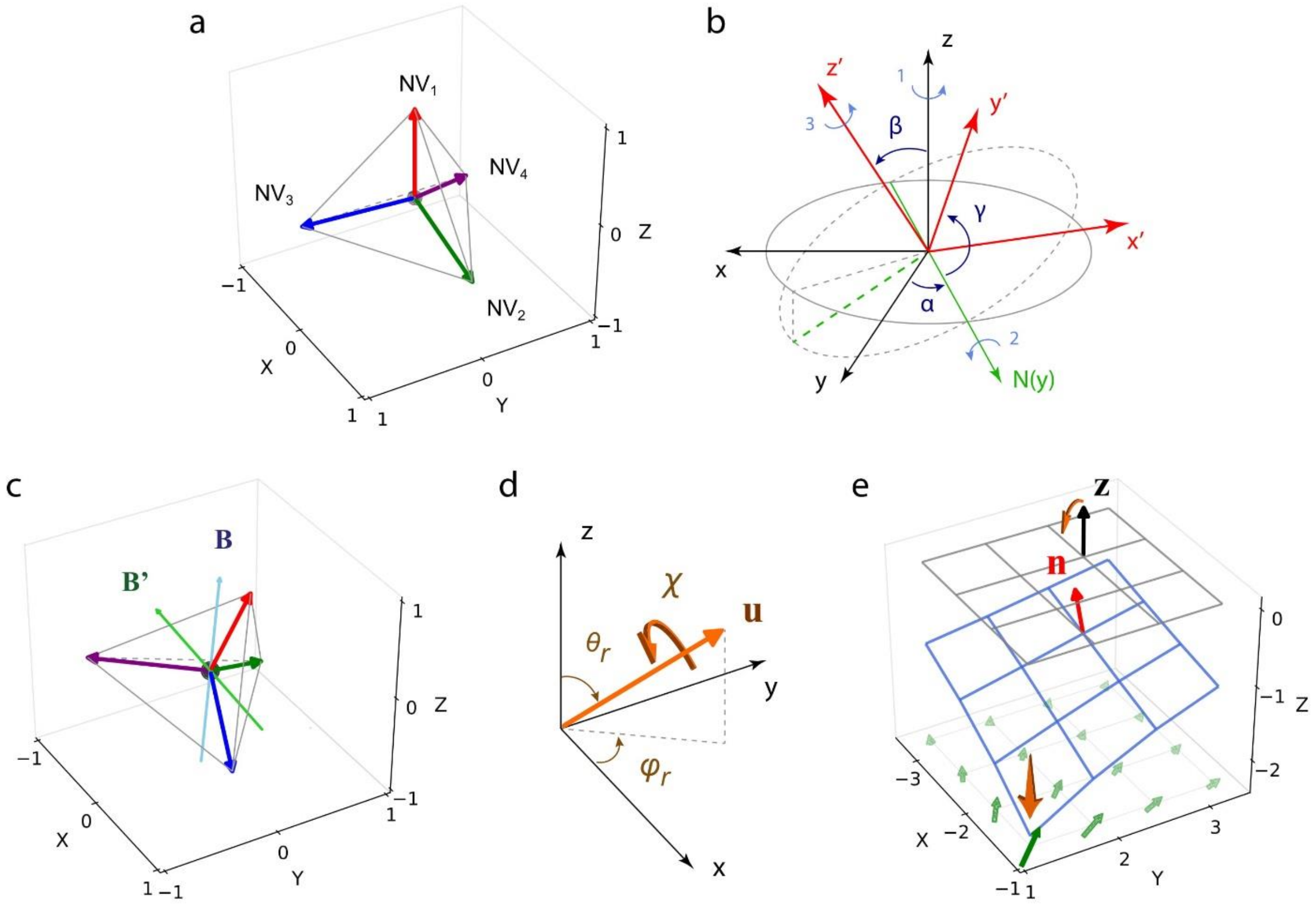

**Fig. S2** | Geometry of the ND rotation measurement. **a,** The four crystallographic orientations of the NV centers in a certain ND (labelled as ND0). The laboratory coordinates are chosen such that $(111)$ direction ($\mathbf{NV}_1^0$) is along the *z*-axis of and the $(1\bar{1}\bar{1})$ direction ($\mathbf{NV}_2^0$) is in the *x*-*z* plane. **b,** The Euler angles representing the orientation of an arbitrary ND, obtained from the orientation of ND0 by three rotations in sequence (1, 2, and 3). **c,** The two magnetic fields and the orientation of the ND used in the simulations. **d,** An rotation with the rotation axis (orange arrow) defined by ($\theta_\mathrm{r}$,$\phi_\mathrm{r}$) and the rotation angle $\chi$. **d,** Illustration of the sample surface with (blue grid) and without (grey grid) indentation. The black and red arrows are the normal vectors of the surfaces. The green arrows are the surface gradient field.

$$\mathbf{R}_\mathrm{I}(\alpha,\beta,\gamma)=\mathbf{R}_z(\alpha)\mathbf{R}_y(\beta)\mathbf{R}_z(\gamma)=\begin{pmatrix} c_\alpha c_\beta c_\gamma - s_\alpha s_\gamma & -c_\gamma s_\alpha - c_\alpha c_\beta s_\gamma & c_\alpha s_\beta \\ c_\alpha s_\gamma + c_\beta c_\gamma s_\alpha & c_\alpha c_\gamma - c_\beta s_\alpha s_\gamma & s_\alpha s_\beta \\ -c_\gamma s_\beta & s_\beta s_\gamma & c_\beta \end{pmatrix}, \quad \text{(S2-2)}$$

where $\mathbf{R}_z$ and $\mathbf{R}_y$ are the matrices of the rotations about the $z$-, and $y$-axes, respectively, and $c$ and $s$ stand for Cosine and Sine with $c_x=\cos x$ and $s_x=\sin x$. The corresponding NV orientations in ND-I are related to those of ND0 by $\mathbf{NV}_i^\mathrm{I}=\mathbf{R}_\mathrm{I}(\alpha,\beta,\gamma)\mathbf{NV}_i^0$. Figure S2c shows the orientation of an ND with $(\alpha,\beta,\gamma)=(45°,20°,65°)$.

The rotation of an ND can be described by the rotation axis $(\theta_\mathrm{r}, \phi_\mathrm{r})$ (in the spherical coordinates) and the rotation angle $\chi$ (see Fig. S2d). The rotation matrix $\mathbf{R}$ is defined by[2]

$$\mathbf{R}(\theta_\mathrm{r}, \phi_\mathrm{r}, \chi)\mathbf{v} = (\cos\chi)\mathbf{v} + (\sin\chi)\mathbf{u}\times\mathbf{v} + (1-\cos\chi)\mathbf{u}\mathbf{u}\cdot\mathbf{v}, \tag{S2-3}$$

where $\mathbf{v}$ is an arbitrary vector, and $\mathbf{u} = (\sin\theta_\mathrm{r}\cos\phi_\mathrm{r}, \sin\theta_\mathrm{r}\cos\phi_\mathrm{r}, \cos\theta_\mathrm{r})$ is the unit vector of the rotation axis. The orientation of an ND after the rotation is characterized by the Euler angles $(\alpha', \beta', \gamma')$. The rotation $\mathbf{R}(\theta_\mathrm{r}, \phi_\mathrm{r}, \chi)$ of the ND was obtained by comparing its Euler angles before and after the operation, that is $\mathbf{R}(\theta_\mathrm{r}, \phi_\mathrm{r}, \chi)\mathbf{R}_\mathrm{I}(\alpha, \beta, \gamma) = \mathbf{R}_\mathrm{I}(\alpha', \beta', \gamma')$.

## (2) Determination of the NV orientations

In the presence of known magnetic field(s) (calibration of the magnetic fields can be found in SI Note 3), the orientations of NV centres can be determined from the experimentally measured frequency shift of the ODMR[3] spectra. The ground state of an NV center is a spin triplet state ($S = 1$) with a zero-field splitting $D \approx 2.87$ GHz between the $m_S = 0$ and $m_S = \pm 1$ states defined along the NV axis. The Hamiltonian of the NV center is written as[4]

$$H_i = D\left(\mathbf{NV}_i \cdot \hat{\mathbf{S}}\right)^2 + \gamma_e \mathbf{B}\cdot\hat{\mathbf{S}}, \tag{S2-4}$$

in the presence of a magnetic field $\mathbf{B}$, where $\hat{\mathbf{S}} = (\hat{S}_x, \hat{S}_y, \hat{S}_z)$ is the spin-1 operator and $\gamma_\mathrm{e} \approx 28\ \mathrm{MHz\ mT^{-1}}$ is the electron gyromagnetic ratio. By diagonalizing the Hamiltonian, the transition frequencies $f_i^{\pm}$, up to the second order perturbation, are determined as[3]

$$f_i^{\pm} = D + \frac{3\gamma_\mathrm{e}^2 B^2}{2D}\sin^2\psi_i \pm \gamma_\mathrm{e} B\cos\psi_i\sqrt{1 + \frac{\gamma_\mathrm{e}^2 B^2}{4D^2}\tan^2\psi_i \sin^2\psi_i}, \tag{S2-5}$$

where $\psi_i = \cos^{-1}(\mathbf{NV}_i \cdot \mathbf{B}/B)$ is the angle between the NV axis and the external magnetic field $\mathbf{B}$. For a given magnetic field, the orientation of the $\mathbf{NV}_i$ (and hence that of the ND) is determined from the angles $\psi_i$. However, if the ND it is rotated by any angle about the magnetic field **B**, the ODMR frequencies are unchanged. This ambiguity intermingling the orientation can be removed by introducing another magnetic field $\mathbf{B}'$ along a different direction from **B**. In this way, another set of angles $\psi_i'$ can be obtained, which enables the unambiguous determination of the orientation of the ND. By the least-square fitting of the respective transition frequencies $f_i^{\pm}$ (with **B**) and $f_i'^{\pm}$ (with **B'**), with the Euler angles $(\alpha, \beta, \gamma)$ and the zero-field splitting $D$ as the fitting parameters, we obtained the orientation of the ND. Once the ND orientation (characterized by the Euler angles) are determined, the comparison of the orientations before and after the rotation gives the rotation data $(\theta_\mathrm{r}, \phi_\mathrm{r}, \chi)$.

In practice, instead of fitting the transition frequencies, we directly fit the four ODMR spectra under magnetic field **B** or $\mathbf{B}'$ before and after the rotation to deduce the rotation of the ND. The normalized ODMR spectrum is written as[4]

$$S(f) = b - \sum_i C_i \left[\frac{\Delta f^2}{4\left(f - f_i^+\right)^2 + \Delta f^2} + \frac{\Delta f^2}{4\left(f - f_i^-\right)^2 + \Delta f^2}\right], \tag{S2-6}$$

where $b$ is the baseline, $C_i$ represents the ODMR contrast of the NV centres along the $i$th direction and $\Delta f$ is the linewidth (FWHM). In the least-square fitting, the initial orientation of the ND were firstly deduced from the two ODMR spectra before the rotation by employing the Euler angles $(\alpha, \beta, \gamma)$, the baselines, the contrasts, the FWHM, and the shift of the zero-field splitting $\Delta D$ (slightly varied among NDs) as the fitting parameters. Then, the rotation of the ND were determined by fitting the two ODMR spectra after the rotation with the specific rotation axis and angle $(\theta_\mathrm{r}, \phi_\mathrm{r}, \chi)$, the contrasts, and the baselines as fitting parameters. The Euler angles, the FWHM, and $\Delta D$ were fixed to be the results of the before rotation fitting. The fitting results of the contrasts and the baselines were approximately identical for the ODMR measurements of the same

ND. The slightly variation of the parameters were due to the laser and the MW power fluctuation between the ODMR measurements.

### (3) Reconstruction of deformation from ND rotation data

The surface of the material before indentation is assumed to be horizonal ($z = 0$) with the surface normal $\hat{\mathbf{z}} = (0, 0, 1)$ (Fig. S2d). The deformed surface due to the AFM indentation becomes $z(\boldsymbol{\rho})$, where $\boldsymbol{\rho}$ is a point on the $x$-$y$ plane. The normal vector of the deformed surface is defined as $\mathbf{n} = (-g_x, -g_y, 1)$, where $\mathbf{g}(\boldsymbol{\rho}) = (g_x, g_y) = \boldsymbol{\nabla}_{\boldsymbol{\rho}} z(\boldsymbol{\rho})$ is the gradient field of the deformed surface. If there is neither translation nor rotation relative to the surface of a given ND docked at point $\boldsymbol{\rho}$, the unit normal vector of the deformed surface can be derived from the rotation data of the ND $\mathbf{R}(\theta_{\mathrm{r}}, \phi_{\mathrm{r}}, \chi)$ from

$$\hat{\mathbf{n}}(\boldsymbol{\rho}) = (n_x, n_y, n_z) = \mathbf{R}(\theta_{\mathrm{r}}, \phi_{\mathrm{r}}, \chi)\,\hat{\mathbf{z}}. \tag{S2-7}$$

In turn, the gradient field $\mathbf{g}(\boldsymbol{\rho})$ is $\mathbf{g}(\boldsymbol{\rho}) = \left(-\frac{\hat{n}_x}{\hat{n}_z}, -\frac{\hat{n}_y}{\hat{n}_z}\right)$. The deformed surface $z(\boldsymbol{\rho})$ was reconstructed from $\mathbf{g}(\boldsymbol{\rho}_n)$ at a discrete set of locations $\{\boldsymbol{\rho}_n\}$, by minimizing the global least-square cost function[5]

$$\epsilon[z(\boldsymbol{\rho})] = \sum_n |\mathbf{g}(\boldsymbol{\rho}_n) - \boldsymbol{\nabla}_{\boldsymbol{\rho}} z(\boldsymbol{\rho}_n)|^2. \tag{S2-8}$$

### (4) Deformation and ODMR simulation

The axisymmetric deformation $z(\rho)$ of a homogenous material caused by the indentation of an AFM tip (cone-shaped with tip radius 100 nm and half-angle 12°) was numerically simulated with a linear elastic model (the Hertzian model)[6,7]. The result is shown as the blue line in Fig. 1c in the main text. For an ND docked on the deformed surface, the corresponding rotation angle $\chi$ of the ND induced by the indentation was calculated as $\chi(\rho) = \tan^{-1}[\partial_\rho z(\rho)]$ and the result is shown as the green line in Fig. 1c.

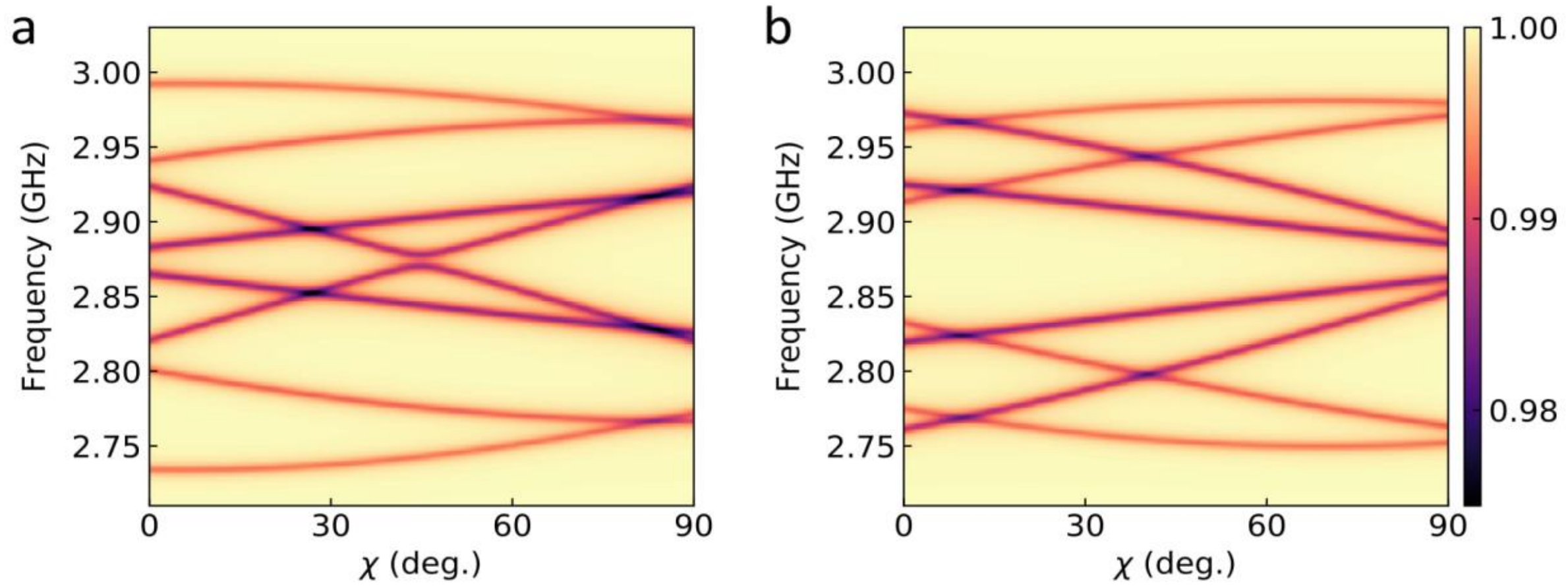

**Fig.S3** | Simulated ODMR spectra as a function of the rotation angle $\chi$ under magnetic fields **B** (**a**) and **B**′ (**b**).

Taking an ND with initial orientation specified by the Euler angles $(\alpha, \beta, \gamma) = (45°, 20°, 65°)$ (configuration shown in Fig. S2c) and rotated about the axis specified by $(\theta_\mathrm{r}, \phi_\mathrm{r}) = (90°, 25°)$, we simulated the ODMR spectra of the ND as a function of the rotation angle $\chi$, for two different magnetic fields $\mathbf{B} = (60\ \mathrm{Gauss}, 5°, 90°)$ and $\mathbf{B}' = (60\ \mathrm{Gauss}, 30°, 270°)$ (shown in Fig. S2c). The contrasts of the NV centers were set to $\{C_i\} = (0.7\%, 1.2\%, 0.8\%, 1.4\%)$ and $\Delta f = 7$ MHz in all the cases. The contrasts and linewidths of the ODMR spectra with the two magnetic fields were set to be the same due to the small angle (35°) between the two fields. Two examples of such ODMR spectra (with the two rotation angles $\chi = 2°$ and $8°$) are presented in Fig. 1c in the main text, and the frequency shifts in all simulated ODMR spectra are plotted in Fig. S3.

**Supplementary Information Note 3. Method validation**

Feasibility of the proposal was firstly evaluated by measuring the rotation of a bulk diamond (containing NV ensembles) of known orientations. Figure S4a shows an optical image of the bulk diamond. We set the $(0\bar{1}1)$ and $(011)$ axes of the diamond as the $x$- and $y$-axes, respectively. The four directions of the NV centers are shown in Fig. S4b with color arrows. The corresponding Euler angles of this diamond were $(\alpha, \beta, \gamma) = (0°, 125.3°, 0°)$. The applied magnetic field **B** (light blue arrow in Fig. S4b) had two components with one from a permanent magnet and the other from an electromagnet placed in the vicinity of the sample. $\mathbf{B}'$ (green arrow in Fig. S4b) was obtained simply by inverting current flow direction in the electromagnet. Calibration of **B** and $\mathbf{B}'$ was carried out by fitting the ODMR spectra (black dotted lines in Fig. S4c) taken from the bulk diamond (with the above specified Euler angles) under the magnetic field **B** or $\mathbf{B}'$. Least-square fitting was adopted (Eq. (S2-6)), with the magnetic field, the shift of the zero-field splitting $\Delta D$, the baselines $b$, the contrasts $C_i$ and the linewidth $\Delta f$ being the fitting parameters. The two magnetic fields were calibrated as $\mathbf{B} = (51.2\ \text{Gauss}, 84.2°, 169.7°)$ and $\mathbf{B}' = (54.5\ \text{Gauss}, 70.3°, 169.9°)$. The fitting results of the ODMR spectra are shown in grey lines in Fig. S4c. The other fitting parameters were obtained as $\Delta D = 1.1$ MHz , $\{C_i\} = (1.6\%, 1.4\%, 1.1\%, 1.5\%)$, and $\Delta f = 6.4$ MHz (broadened by the $^{14}$N hyperfine structure of the ODMR spectrum).

The bulk diamond sample was rotated anticlockwise about the *z*-axis. The optical image of the bulk diamond after the rotation is shown in the lower panel of Fig. S4a. Two ODMR spectra under **B** or $\mathbf{B}'$ were acquired after the rotation (red dotted lines in Fig. S4c). By fitting the ODMR spectra with the known initial Euler angles of the diamond sample and the pre-calibrated magnetic fields, the rotation axis and angle were deduced as $(\theta_{\rm r}, \phi_{\rm r}, \chi) = (4°, 290°, 20.8°)$. The $\theta_{\rm r} \sim 4°$ suggesting a small deviation of the rotation axis (grey arrow plotted in Fig. S4b) from the *z*-axis ($\theta_{\rm r} = 0°$). It was possibly caused by the manual manipulation of the sample stage for rotation operation. The deduced rotation angle $\chi \sim 20.8°$ was consistent with the rotation angle ($\sim 21°$) measured from the optical images before and after the sample rotation (Fig. S4a).

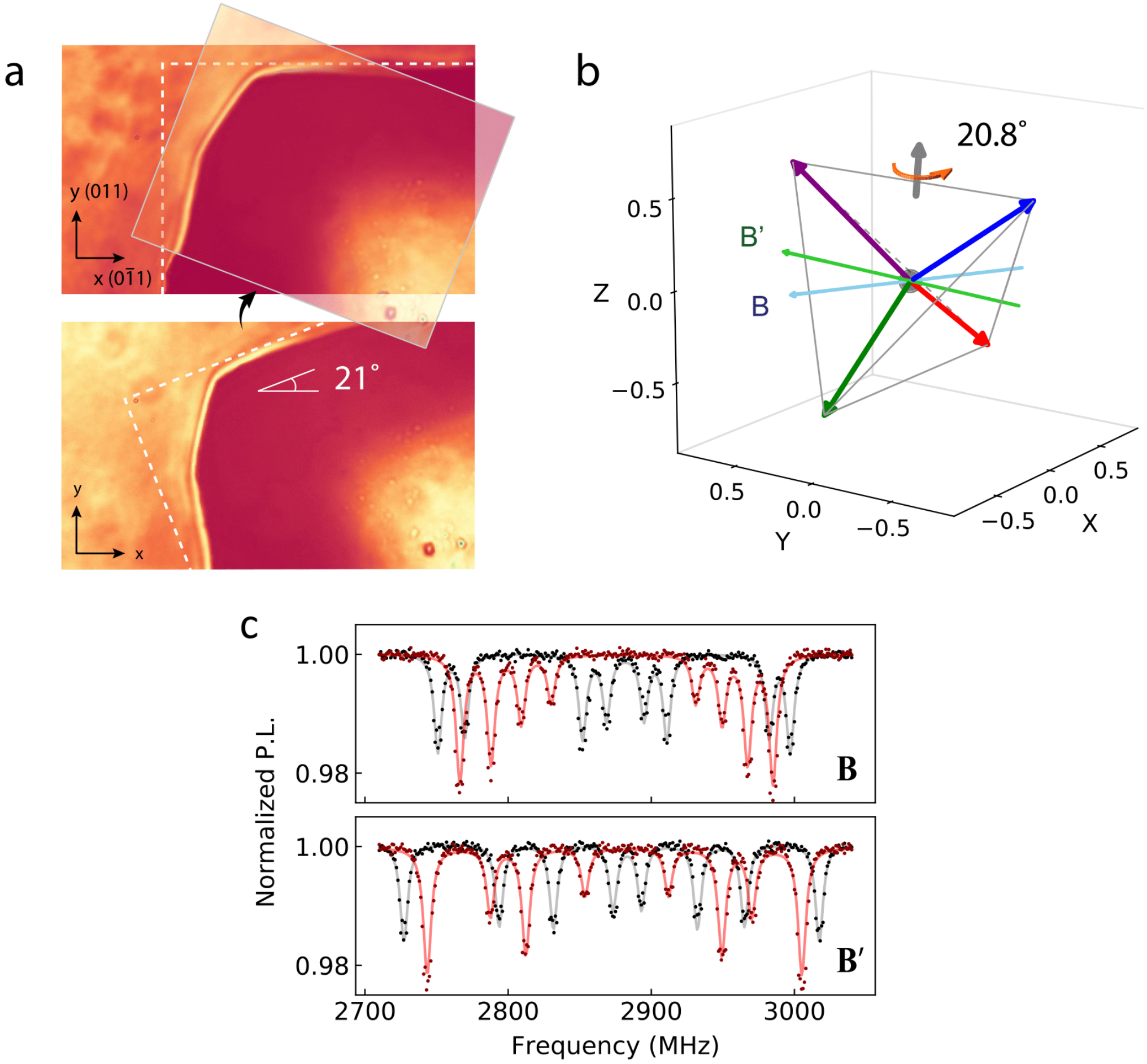

**Fig. S4** | Rotation measurement of a bulk diamond. **a,** Optical image of the bulk diamond before (upper) and after (bottom) the rotation. Before rotation, the sample was positioned with the $(0\bar{1}1)$ and $(011)$ crystallographic axes along the *x*-, and *y*-axes of the laboratory frame, respectively. A 21° clockwise rotation is applied to the image after rotation in order to regain its original orientation. **b,** Schematic showing the configuration of the magnetic fields **B** and **B′**, and the orientations of the NV centers inside the bulk diamond. The grey arrow marks the rotation axis. **c,** ODMR spectra of the bulk diamond before (black dots) and after (red dots) rotation under the magnetic field **B** (upper) or **B′** (bottom). The grey and red lines are the best fitting.

## Supplementary Information Note 4. Reconstructing the deformation of a PDMS thin film

### (1) Fabrication of PDMS films and their characterizations

PDMS was synthesized by mixing the precursor and cross linker with a ratio of 5:1. A microwave antenna was amounted a glass slide. The PDMS was then spin-coated onto the glass slide. After degassing, the surface of the PDMS film was modified by $O_2$ plasma to generate a surface oxidized layer[8,9]. At last, the NDs were spin coated on the PDMS film. The thickness of the PDMS film was measured to be $\sim 50$ µm (from optical measurement). To estimate the thickness of the oxidized surface layers, we prepared a cross-sectional PDMS sample by putting together two PDMS films with oxidized surfaces face to face. Figure S5a shows the AFM stiffness mapping of the cross-sectional PDMS. The dark region represents the bulk PDMS, and the bright region is the surface layer (with larger stiffness). A $\sim 500$ nm thickness of the surface oxidation layer was estimated from the thickness between the two dashed lines. AFM nanoindentation was performed on the bulk PDMS far from the oxidized surface, and Young's modulus of the bulk PDMS was obtained as $E \sim 0.78$ MPa with the Hertzian model[6,7].

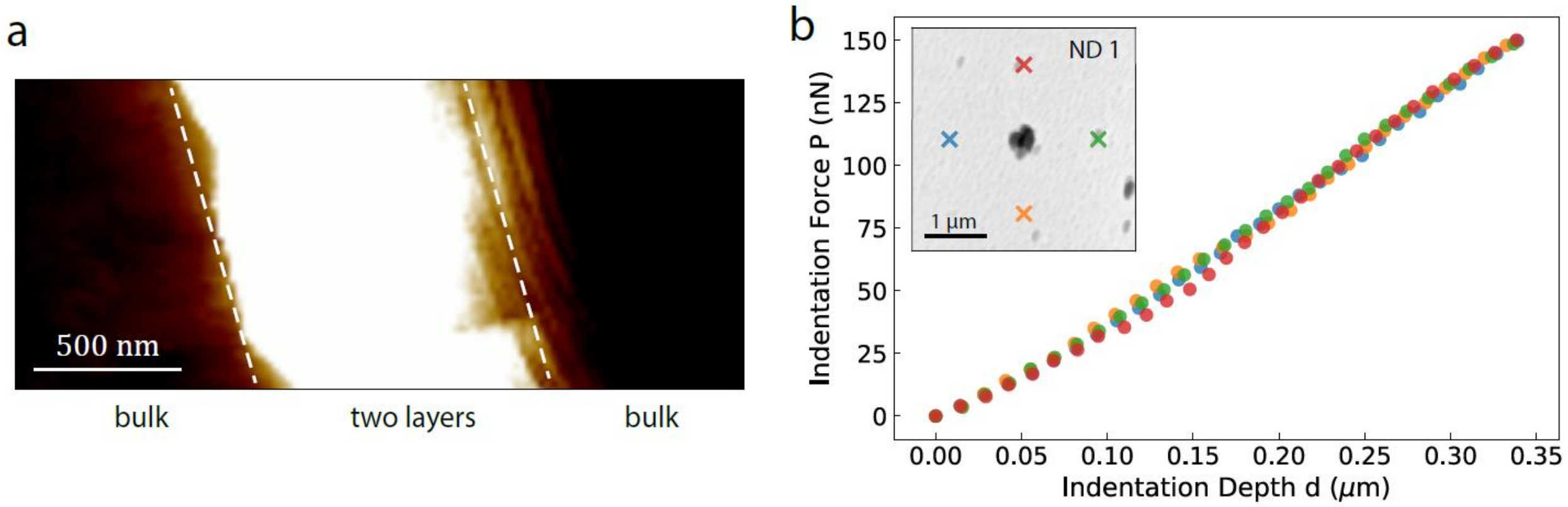

**Fig. S5** | Mechanical properties of oxidized PDMS film. **a,** The AFM stiffness mapping of a cross-sectional PDMS sample with surface oxidization. The white dashed lines separate the bulk and the surface layers. **b,** Force-depth profiles of the nanoindentations at different spots of the PDMS film. The inset shows the AFM image of the PDMS film, and the color crosses label the spots where the nanoindentations were applied.

A silicon nitride cantilever (DNP10-A Bruker) was applied to image and indent the PDMS film. After calibrating the spring constant and deflection sensitivity of this tip, the PDMS film was imaged in a peak force tapping mode and then indented in a ramp mode. Figure S5b shows a few typical load-depth profiles obtained by nanoindentation performed at the spots labeled in the inset. The almost identical load-depth profiles of indentations carried out at different location on the sample suggest the homogeneity in the *x*-*y* plane of the PDMS film.

## (2) Methods of orientation measurements of surface anchored NDs under an AFM tip indentation

NDs were dispersed on the surface of PDMS film by spin coating. In this protocol, dilute ND solution (2 µg $mL^{-1}$) was employed to avoid NDs aggregation on the surface. Figure S6a shows the AFM topography image with NDs located on a 40 × 40 µm PDMS surface, correlating well with the confocal image (Fig. S6b) taken from the same area by the pattern of ND spatial distribution (marked by the white dash rectangles). The indentation experiment was carried out by lowing the AFM tip into the PDMS film in the proximity of a chosen ND (e.g. point 30 as shown in Fig. 2a in the main text) at a constant force $P = 150$ nN. ODMR spectra of the ND were recorded with and without an indentation at two pre-calibrated specific magnetic fields (using bulk diamond, see SI Note 3) as $\mathbf{B} = (69.6\ \text{Gauss}, 89.4°, 0.0°)$ or $\mathbf{B}' = (75.5\ \text{Gauss}, 104.3°, 13.6°)$.

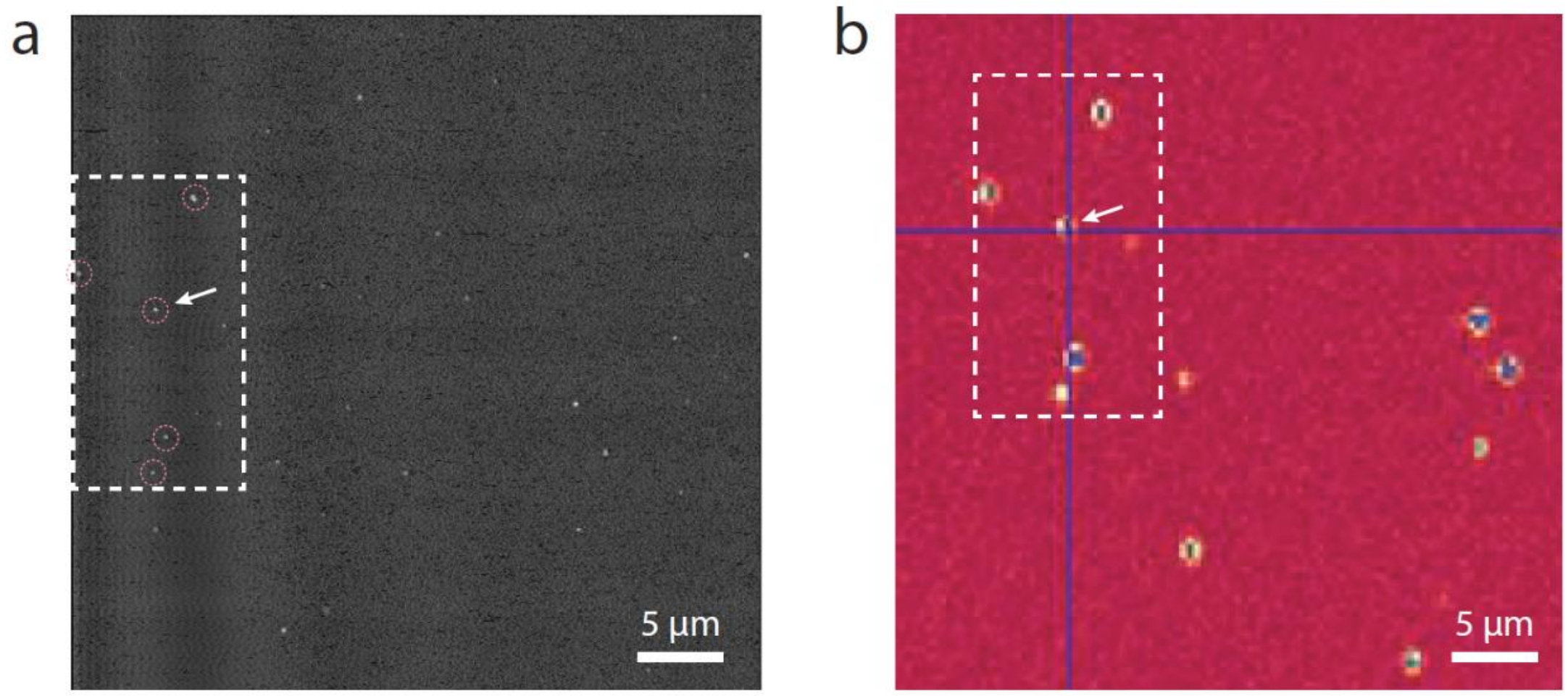

**Fig. S6** | Correlated AFM and fluorescence images. **a,** AFM topography image with NDs (white dots) on a 40 × 40 μm PDMS surface. Some NDs are circled by pink dash lines. **b,** Confocal image taken from roughly the same region. The two images are correlated (marked by white dash rectangles) by the pattern of the NDs distribution. The white arrows indicate the same ND in the two images.

Using the methods described in SI Note 1-2, the initial orientation of the ND (without indentation) was deduced from its ODMR spectra under $\mathbf{B}$ and $\mathbf{B}'$ and the corresponding fittings (black dots and lines in Fig. S7). The characteristic Euler angles were determined as $(\alpha, \beta, \gamma) = (-156.2°, 85.1°, 71.2°)$. Rotation of the specific ND was then determined by fitting the ODMR spectra with the indentation as $(\theta_{\mathrm{r}}, \phi_{\mathrm{r}}, \chi) = (90°, 76°, 2.7°)$ (red dots and lines in Fig. S7).

### (3) Deformation reconstruction

For the *x*-*y* plane homogeneous medium, we adopt the single-ND method to reconstruct the deformation. Experimentally, the ND orientation sensor was fixed on a position with an AFM tip performing nanoindentation around the ND. This single-ND method is possible because on the homogeneous material the deformation at position A upon an indentation at position B should be the same as the deformation at B upon an indentation at A. The AFM indentation was performed in the proximity of the ND at the centre (0,0) in a programmed manner (indentation locations marked as red dots in Fig. 2a in the main text, the sequence of the indentations from 1-92 is marked by the blue arrows). The rotation of the ND upon each indentation were derived from the ODMR spectra for various displacements $\boldsymbol{\rho}$ $(\rho, \theta)$ of the ND from the AFM indentation positions (as shown in Fig. 2b in the main text). The deformation due to the indentation by an AFM tip applied to the origin (0,0) was reconstructed from the gradient field $\mathbf{g}(\boldsymbol{\rho})$ deduced by the rotation data of the ND (details see SI Note 2-3). In the reconstruction process, the missing gradient field in the center and the edge of the deformed surface was estimated by the curvature-minimizing interpolant[10].

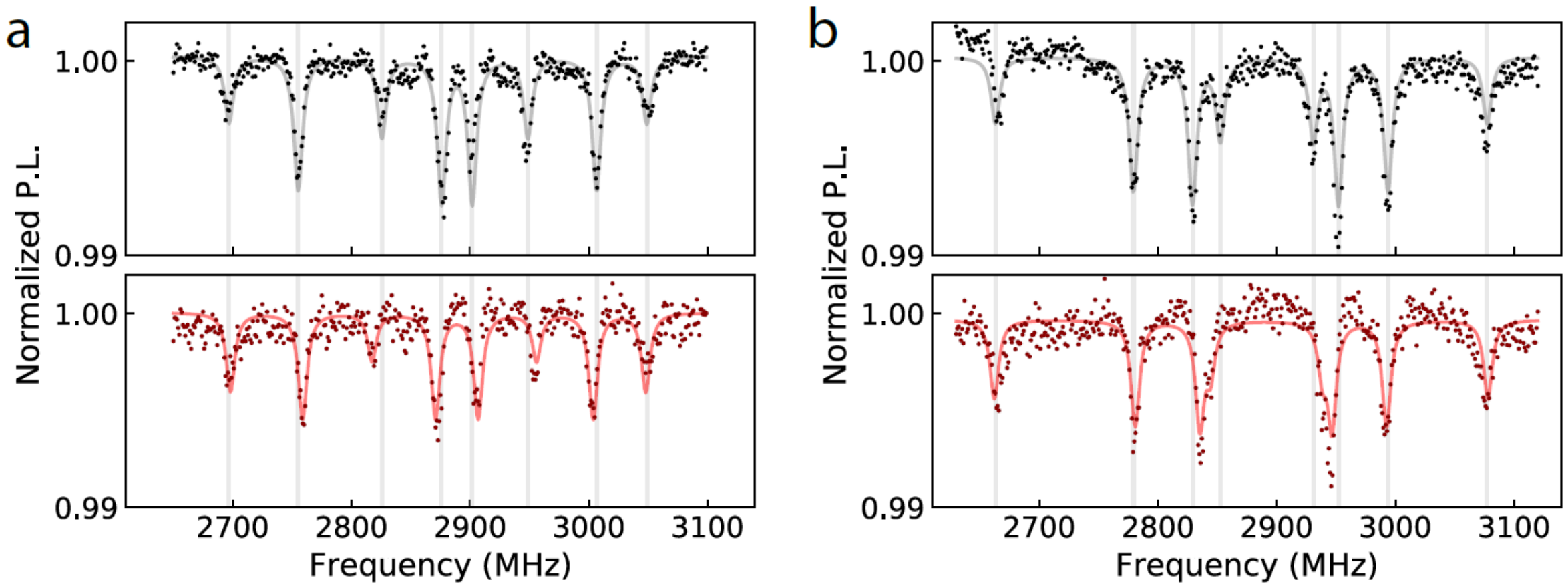

**Fig. S7 |** Typical ODMR spectra with and without an indentation. The ODMR spectra without (black dots) and with (red dots) indentation under magnetic field **B** (**a**) or **B′** (**b**). Black and red solid lines are the fitting results of the ODMR spectra. Grey lines indicate the positions of the resonance peaks before indentation.

The experiments have been repeated using different NDs on different locations of the PDMS film, and similar results were obtained. Here we show results obtained from another NDs as examples (ND2 and ND3). The AFM images of the two NDs are shown in Fig. S8a and b, respectively. The same methodology has been adopted. Initial Euler angles of ND2 and ND3 are $(\alpha, \beta, \gamma) = (-5.0°, 112.7°, 21°)$ and $(-12.1°, 89.6°, -159.0°)$, respectively. By fitting the ODMR spectra, the rotations of the NDs with different indentations were obtained. Similar to data (of ND1) present in the main text (Fig. 2b), the rotation of the ND2 and ND3 for various displacements $\boldsymbol{\rho}\ (\rho, \theta)$ of the ND from the AFM indentation positions are shown in Fig. S8c, and d, respectively, in which the arrows represent the projected rotation axes and the rotation angle.

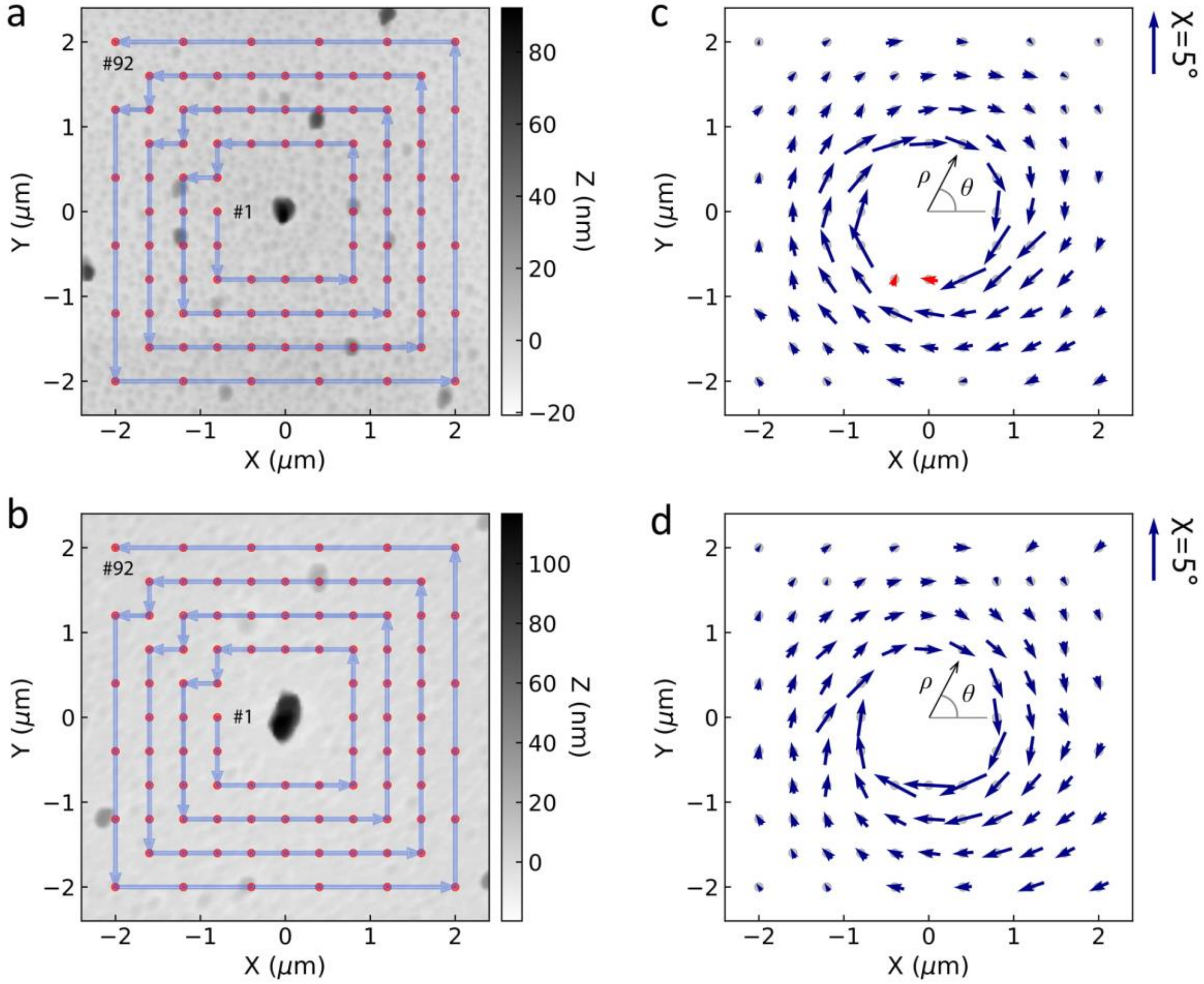

**Fig. S8 |** Deformation reconstruction using the single-ND approach. **a** and **b,** AFM image of typical PDMS surface with the ND2 and ND3 located the center (origin), respectively. The red dots represent the actual indentation locations of the AFM tip and the blue arrows direct the indentation sequence, starting with #1 and ending with #92. **c** and **d,** The rotations of the ND plot as arrows for the displacements of the ND2 and ND3, respectively, from the 92 indentation positions. The size of the arrows represents the magnitude of the rotation angles $\chi$ (scale bar shown on the right), and the direction is the rotation axes projected to the *x*-*y* plane.

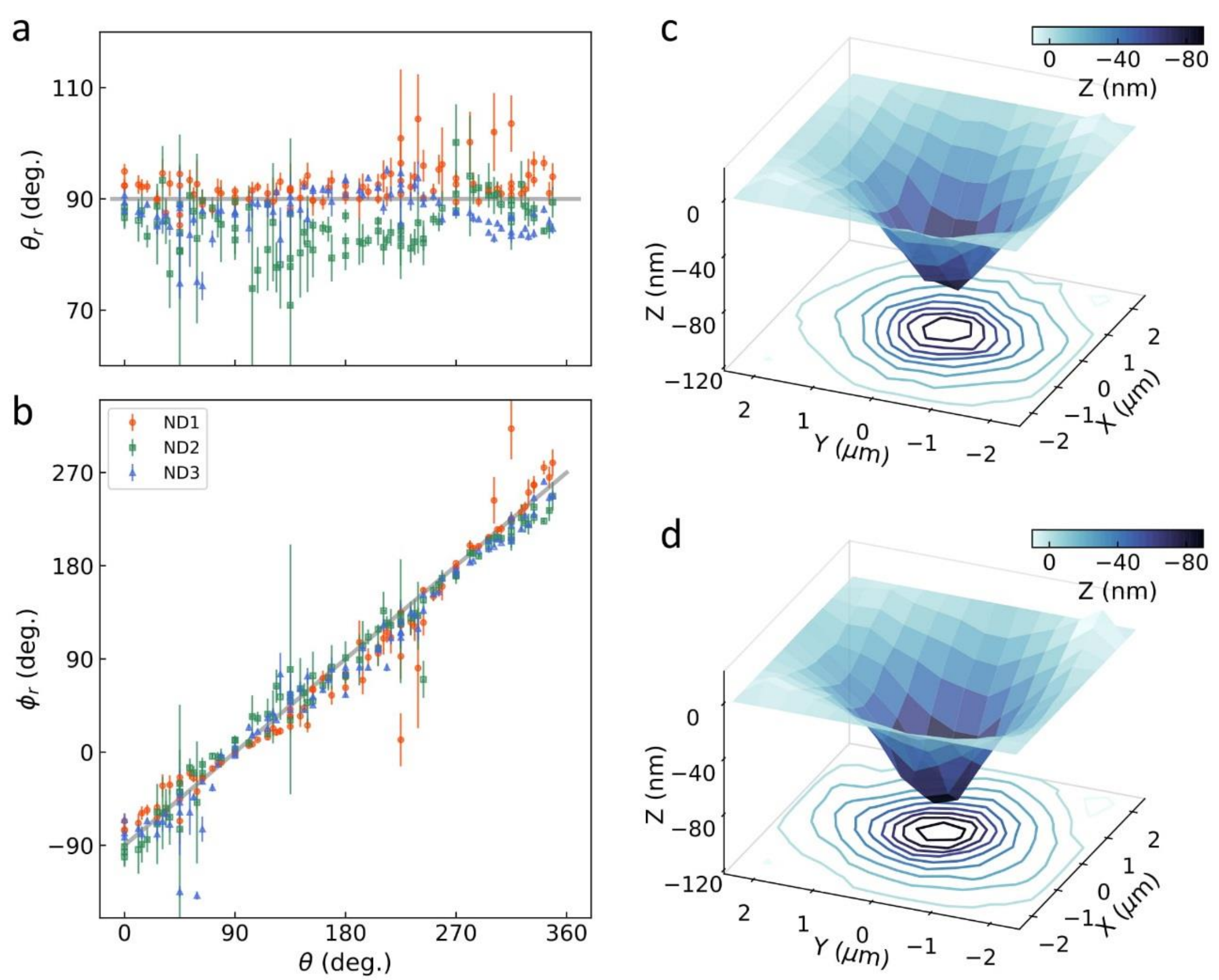

**Fig. S9 |** Deformation reconstruction using the single-ND approach. **a,** The polar angle $\theta_\mathrm{r}$ and **b,** the azimuthal angle $\phi_\mathrm{r}$ of the rotation as functions of $\theta$ for the three NDs. The grey line in **b** is the condition $\phi_\mathrm{r} = \theta - 90°$. **c** and **d,** The reconstructed surface of the PDMS film during the AFM tip indentation for ND2 and ND3, respectively.

The rotation axis is always found on the tangential direction of the circle centered on the origin, suggesting that the NDs always rotates towards to indentation point. The sizes of the arrows represent the rotation angle $\chi$. The larger the separation distance between the indenting location and the NDs, the smaller the rotation angle. Two abnormal points are found in Fig. S8c near the indentation points (marked by red arrows in Fig. S8c). This is likely induced by the tip-sample interactions when the indenting tip is very close to the ND—artificial displacement of NDs can be induced, resulting in the abnormal signals. Together with results obtained from ND1 (presented in the main text), here we plot the polar angle $\theta_\mathrm{r}$ and the azimuthal angle $\phi_\mathrm{r}$ of the rotation axis for all three NDs as a function of $\theta$ in Fig. S9a and b, respectively. The rotation axes of all three NDs are approximately on the *x*-*y* plane (i.e., $\theta_{\mathrm{r}} \sim 90°$ ) and are perpendicular to **ρ** (i.e., $\phi_{\mathrm{r}} \sim \theta - 90°$). The deformed surfaces of ND2 and 3 were reconstructed via the gradient fields as illustrated in Fig. S9c, and d, respectively. The deformed surfaces are approximately axisymmetric about the AFM indentation at (0, 0).

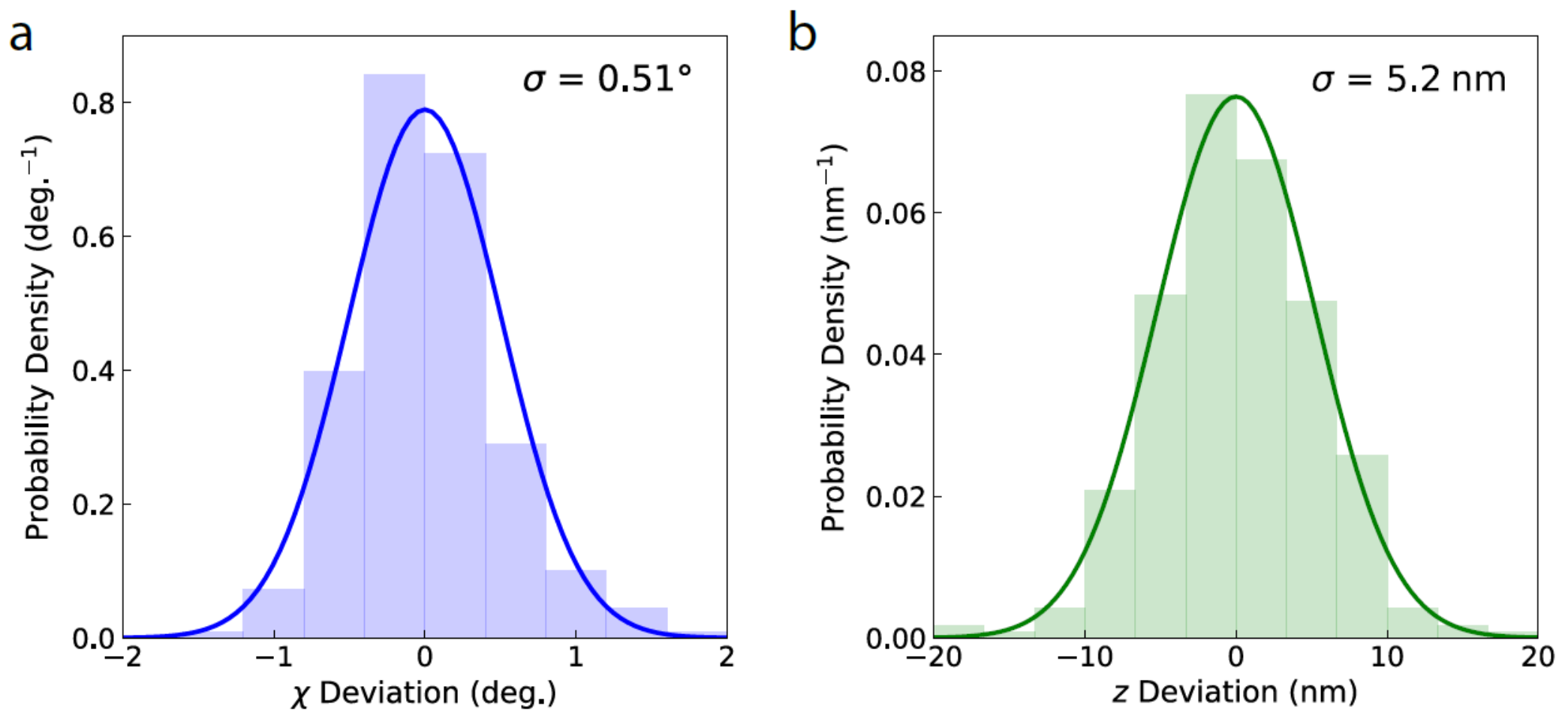

**Fig. S10 |** The probability distribution of the deviation of **a,** the rotation angle $\chi$ and **b,** the $z$-direction deformation.

The deviations of the rotation angle ($\chi$) and the reconstructed surface ($z$) measurement are defined as $\delta\chi = \chi - \langle\chi\rangle_\rho$ and $\delta z = z - \langle z\rangle_\rho$, where $\langle\chi\rangle_\rho$ and $\langle z\rangle_\rho$ are the respective mean value at the same $\rho$ (distance between the ND and the AFM tip indentation) for each ND experiment. Probability distributions of the $\chi$ and $z$ deviations for experimental data taken from all three NDs (blue and green shades) follow Gaussian distribution (blue and green lines) are illustrated in Fig. S10a and b. The standard deviations for the rotation angle and reconstructed surface measurements are about $\sigma_\chi = 0.51°$ and $\sigma_z = 5.2$ nm, respectively. In the present experiments, a sufficiently long measurement time was ensured, so the precision of the deformation measurement is not shot-noise limited, but dominated by the environmental fluctuations, such as system mechanical vibration and temperature fluctuation.

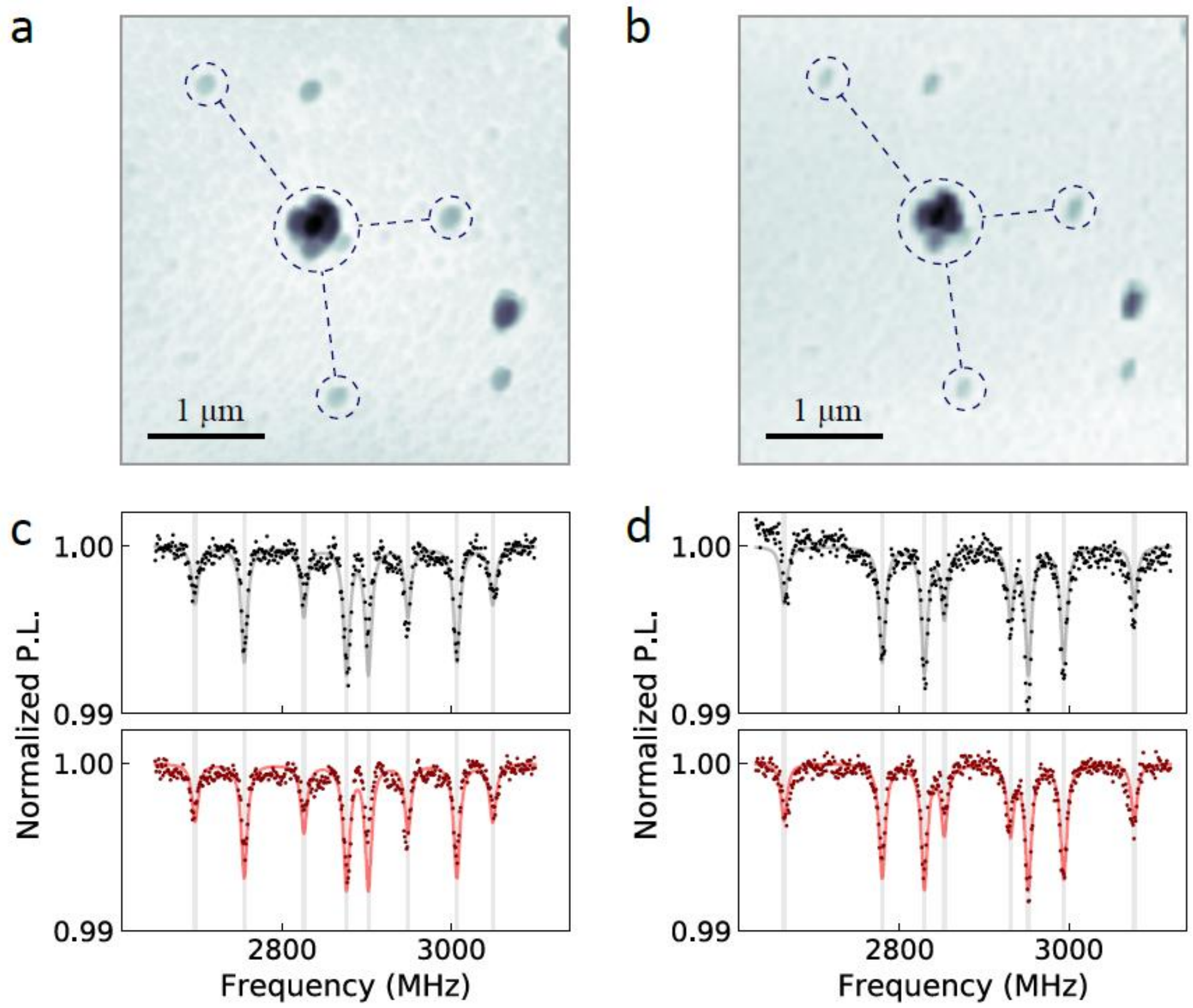

**Fig. S11 |** The position and ODMR spectra of the measured ND before and after the indentation experiments. **a** and **b,** The AFM image of the PDMS surface with the ND1 at center before, and after the indentation experiments, respectively. The patterns drawn in dashed lines matches each other in **a** and **b**. The ODMR spectra before (black dots) and after (red dots) the indentations under magnetic field **B** (**c**) or **B′** (**d**). Black and red lines are the fitting results of the ODMR spectra. Grey lines indicate the positions of the resonance peaks before indentation.

AFM image and the ODMR spectra before and after the indentation experiments are compared for ND1 (Fig. S11). Complete recovery of the PDMS surface after releasing the load is suggested by the absence of surface dent marks (Fig. S11b). Identical patterns on the AFM images before and after the indentations (Fig. S11a and b) reveal the little relative motion of ND particles on such a surface before and after the indentation experiments. Moreover, little change in the orientation of ND1 before and after the indentation was observed, as suggested by the identical resonance frequencies of the corresponding ODMR spectra (Fig. S11c and d). These results suggest that the ND had little relative motion and rotation against the PDMS surface throughout the indentation experiments.

## Supplementary Information Note 5. Simulation of the PDMS deformation.

The indentation induced deformation of PDMS film was simulated using a linear elastic model of a layer/bulk system under an axisymmetric loading, as illustrated in Fig. S12a. The PDMS film (thickness ~50 μm) was modeled as a half-infinite substrate with a surface layer (formed due to oxygen plasma treatment as illustrated in SI Note 4-1). Literature suggested that the oxidized surface layer is stiffer than the bulk PDMS with approximately one-order of magnitude larger elastic modulus[9]. For simplicity, both the surface layer (with thickness $t$, Young's modulus $E_1$, and Poisson's ratio $\nu_1$) and the bulk PDMS (Young's modulus $E_0$ and Poisson's ratio $\nu_0$) were assumed to be isotropic. We adopted the same method as in Ref. 11 to simulated the system. The parameters were chosen as $E_0 = 0.78$ MPa, $E_1 = 9.4$ MPa, $\nu_0 = \nu_1 = 0.33$, and $t = 500$ nm. The shape of the AFM tip was estimated as a cone with tip radius 50 nm and half-angle 25° (from SEM). The indentation force-depth curve was calculated (grey line in Fig. S12b) and agrees well with the experiment results (blue dots in Fig. S12b) obtained by the AFM tip indentation on the PDMS sample. As a comparison, the simulation results obtained by Hertzian model[6,7], assuming the indented material to be a half-infinity homogenous solid with a Young's modulus of $E = 3$ MPa, are presented using dash lines to describe the force-depth profile (Fig. S12b).

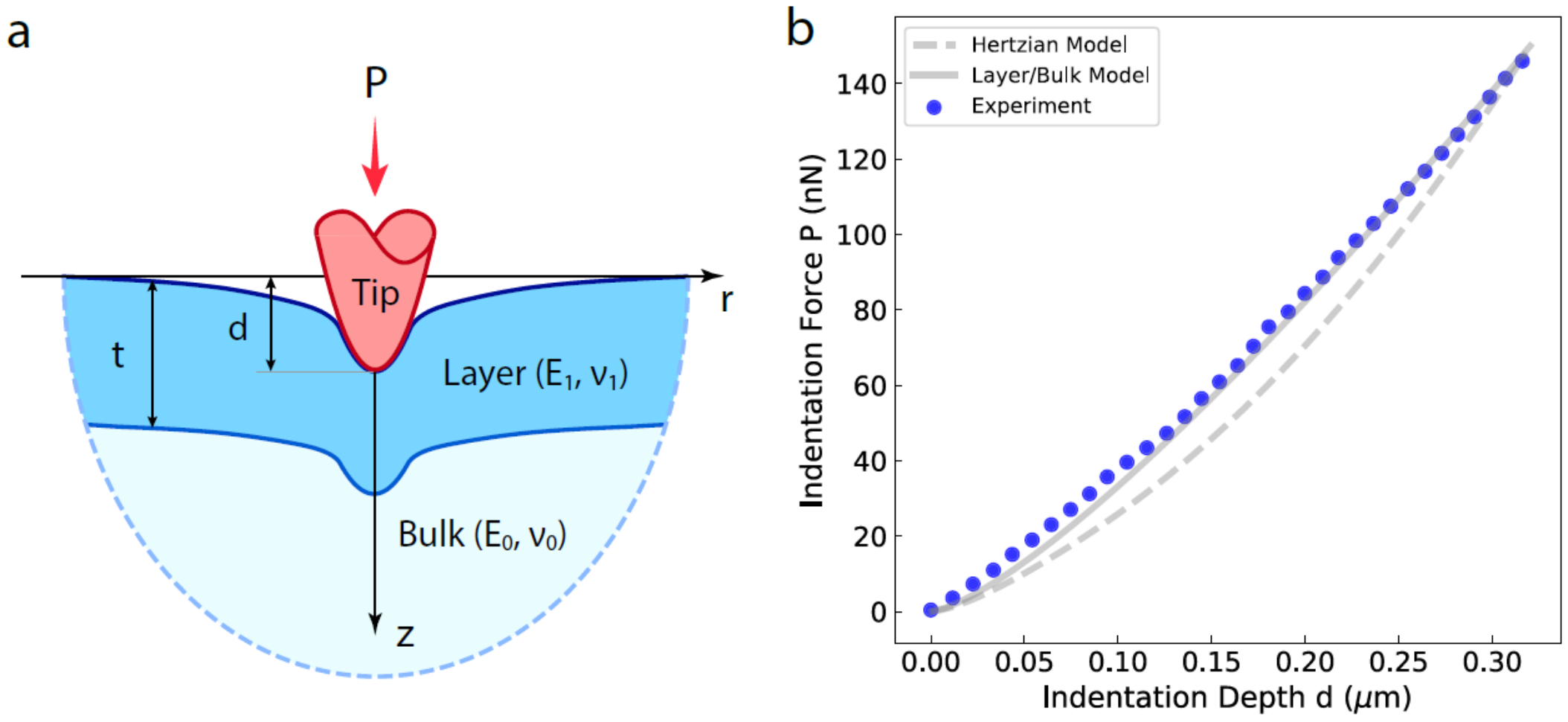

**Fig. S12 |** Simulation of the PDMS deformation. **a,** Schematic of layer/bulk model under indentation. **b,** Indentation force-depth profile of the PDMS film (blue dots). Grey line is the simulation result for the layer/bulk model. Grey dash line is the result for Hertzian model.

## Supplementary Information Note 6. Deformation reconstruction of gelatin upon AFM indentation.

### (1) Fabrication and characterizations of the gelatin particles

In a standard procedure, 1.5 g gelatin was dissolved in 10 mL deionized water at 60 °C. At the same time, 0.2 g Span80 was dissolved in 40 mL toluene in a 100 mL round-bottom flask at 60 °C. Then 2 mL of the previously prepared 0.15 g $mL^{-1}$ gelatin solution was added to the Span80 solution and they were uniformly mixed. After that, the gelatin in toluene emulsion was stirred at 300 rpm and cooled in ambient condition for 1 h. Then, the solidified particles were sonicated for 2 min. Finally, the particles were washed three times with 30 mL acetone, followed by another three times' wash using 30 mL deionized water/time.

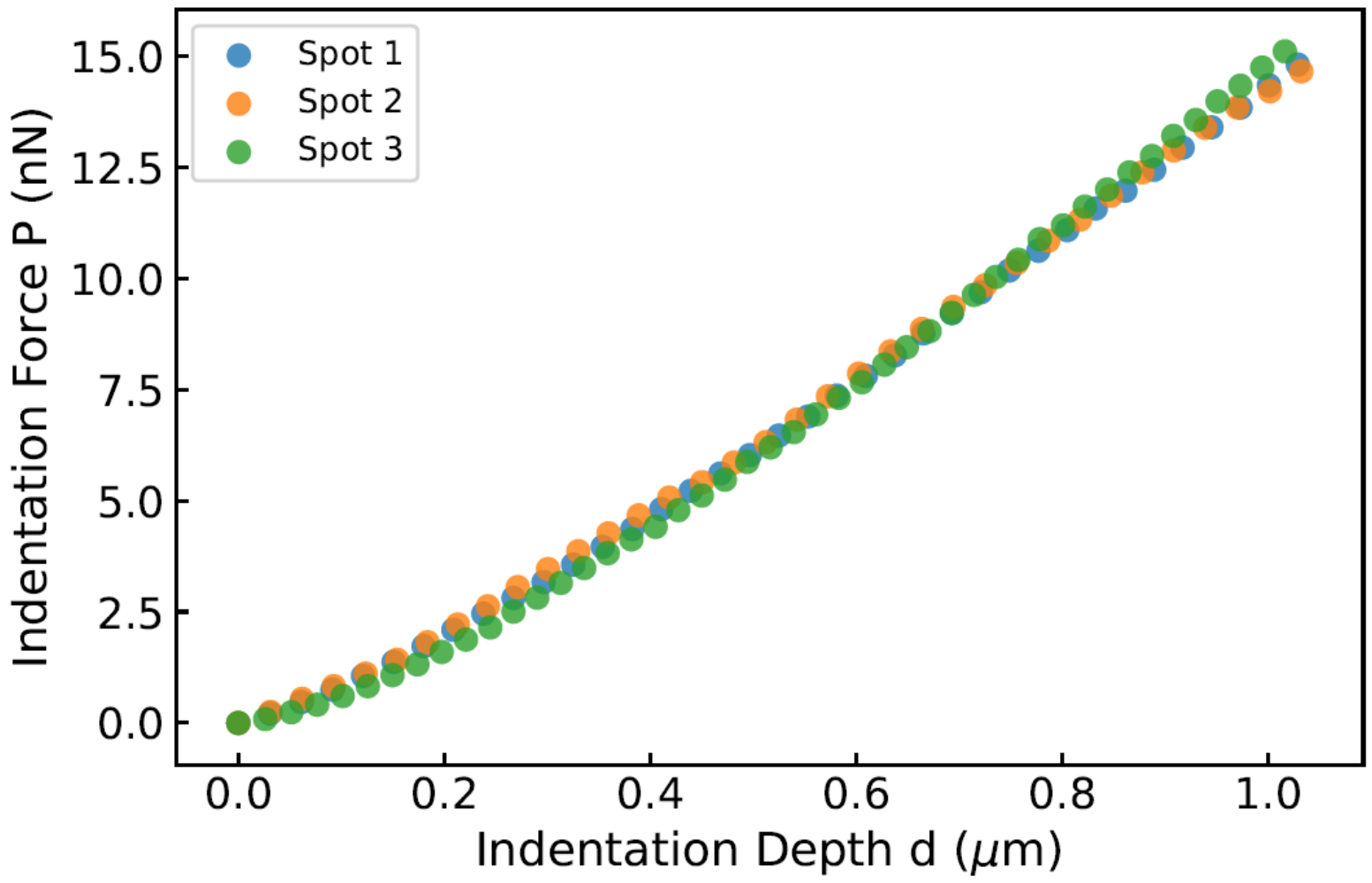

**Fig. S13** | Force-depth profiles of the indentations at different spots on the gelatin particle.

We put the gelatin microparticles on the glass substrate and randomly disperse the NDs on their top. To image and indent soft material in liquid environment, a cantilever (PFQNM-LC, Bruker) with small spring constant of 0.089 N $m^{-1}$ and large tip radius of 0.15 μm was used. We imaged gelatin particles in peak force tapping mode with peak force of 300 pN (Fig. 4b in the main text). Indentation was applied on the gelatin samples at different locations. Figure S13 shows the force-depth plots obtained by indentation performed at different location of the gelatin particle. Similar results were obtained, suggesting homogeneity in *x*-*y* plane of the gelatin particle.

**(2) Orientation measurements of surface anchored NDs under an AFM indentation and deformation reconstruction**

Deformation of gelatin was investigated using multiple NDs dispersed on the sample surfaces. Gelatin particles of 30 μm (diameter) were firstly dispersed on to a culture dish, 20 μL ND aqueous solution (2 μg $mL^{-1}$ concentration) were then dropped on to the gelatin samples. After the NDs were docked on the gelatin samples, water was added to immerse both the sample and AFM tip. The indentation experiments were carried out in the center region of the gelatin particle, where the surface is close to a horizontal plane (the rms roughness is 20 nm with a height difference of ~150 nm from the center to the boundaries of the examined area; the separation distance from the center to the boundary is 2 μm). Upon any specific indentation, rotations of selected NDs at different distance away from the indentation location defines the gradient field there.

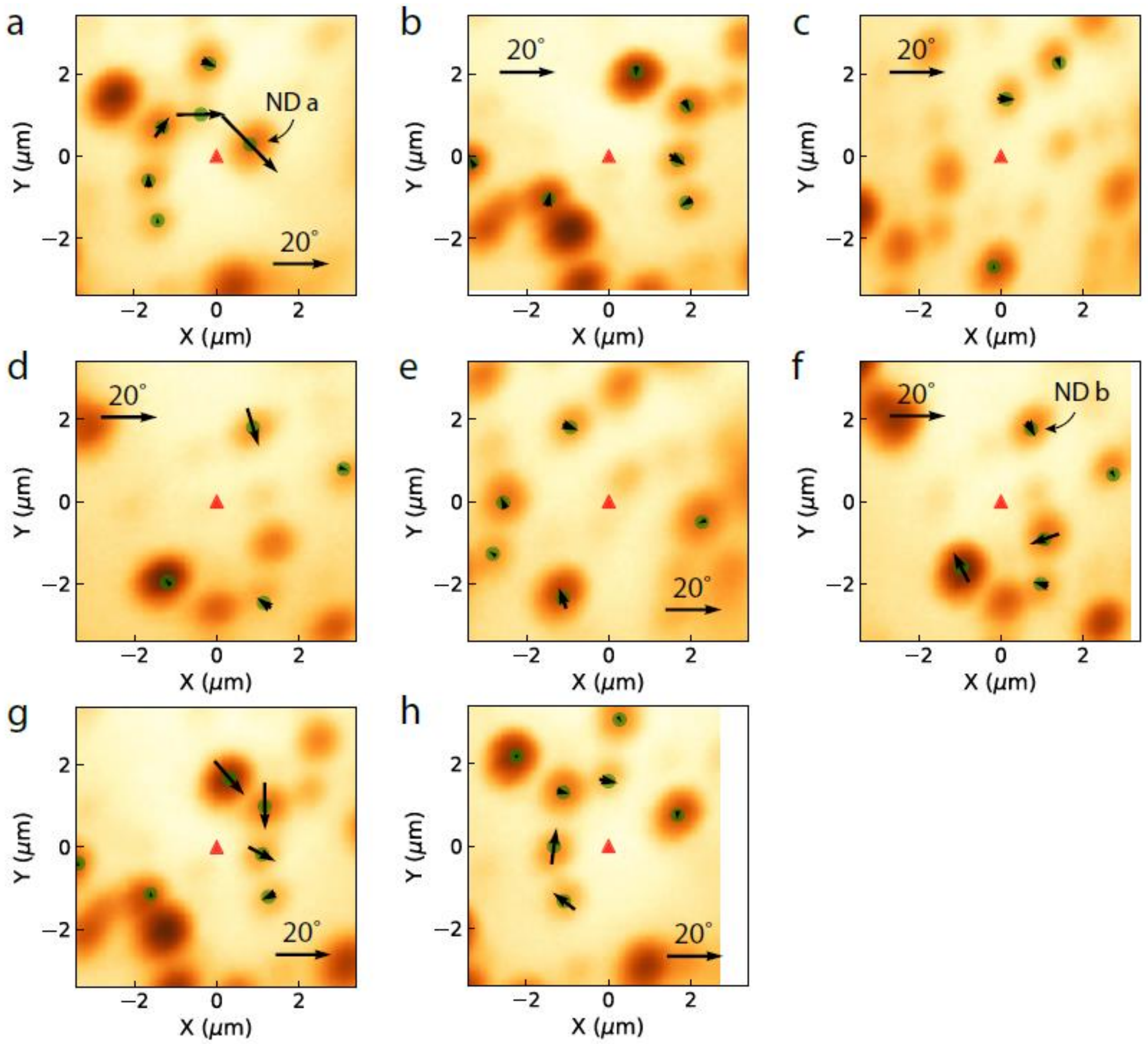

**Fig. S14** | Deformation reconstruction based on the multi-ND approach. **a to h,** Confocal images of gelatin with the indentation carried out at location 1 to 8 (red triangle in the figures). Green circles mark the positions of the NDs. The orientations and the sizes of the black arrows represent the direction of the rotation axes, and the magnitude of the rotation angles of the relevant NDs, respectively.

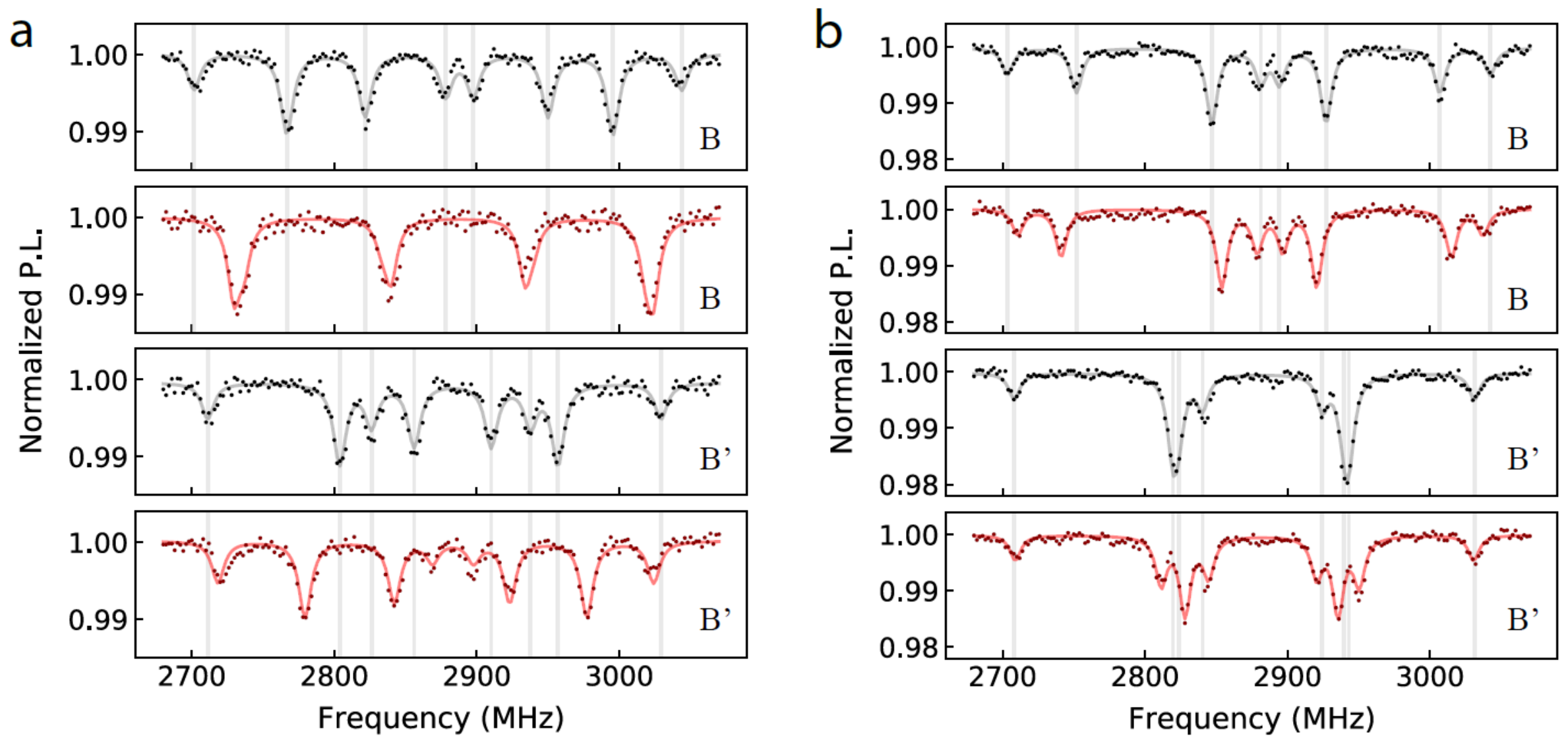

**Fig. S15** | Typical ODMR spectra of the measured NDs with and without indentations. The ODMR spectra without (black dots) and with (red dots) indentation under magnetic fields **B** (upper) or **B**′ (bottom) taken from **a,** NDa (in experiment Set 1 as marked in Fig. S14a) and **b,** NDb (in Set 6 as marked in Fig. S14f). Black and red solid lines are the fitting results of the ODMR spectra. Grey lines indicate the positions of the resonance peaks before indentation.

Eight sets of experiments were carried out on different locations on the gelatin (all within the central region to ensure that the starting planes are horizontal) with identical indentation force $P = 15\ \mathrm{nN}$. The fluorescence images of these 8 sets of experiments are shown in Fig. S14. The indentation location is marked by the red triangle, and the ND locations are marked by green circles. The distance between the indentation location and the measured NDs in each set of the experiments were determined by correlating the respective fluorescence image with the AFM image (for details, see SI Note 1). Pre-calibrated magnetic field $\mathbf{B} = (66.7\ \mathrm{Gauss}, 62.3°, 167.3°)$ or $\mathbf{B}' = (57.8\ \mathrm{Gauss}, 84.8°, 167.3°)$ was applied. Figure S15 gives the ODMR spectra taken from ND a (marked in Fig. S14a) and b (marked in Fig. S14f) with and without indentations as examples. Without indentations, the characteristic Euler angles were determined as $(\alpha, \beta, \gamma)_\mathrm{a} = (156.4°, 83.6°, -60.4°)$ and $(\alpha, \beta, \gamma)_\mathrm{b} = (-16.9°, 93.2°, 172.2°)$ by fitting the ODMR spectra (black dots and lines in Fig. S15). Rotations of the NDs were then determined by fitting the ODMR spectra with indentations, as $(\theta_\mathrm{r}, \phi_\mathrm{r}, \chi)_\mathrm{a} = (95.3°, 314.2°, 28.4°)$ and $(\theta_\mathrm{r}, \phi_\mathrm{r}, \chi)_\mathrm{b} = (110°, 295°, 5.7°)$ (red dots and lines in Fig. S15). The deduced rotation axes and angles of the NDs in each of the indentation experiments are marked in the corresponding locations in Fig. S14. Figure S16a and b plots the polar angles $\theta_\mathrm{r}$ and the azimuthal angles $\phi_\mathrm{r}$ of the rotation axis as functions of the polar angle $\theta$ (ND position in the polar coordinate $(\rho, \theta)$). A reasonable agreement exists between the experimental results and $\phi_\mathrm{r} = \theta - 90°$ (grey line in Fig. S16b), suggesting that the NDs rotate toward the indentation location.

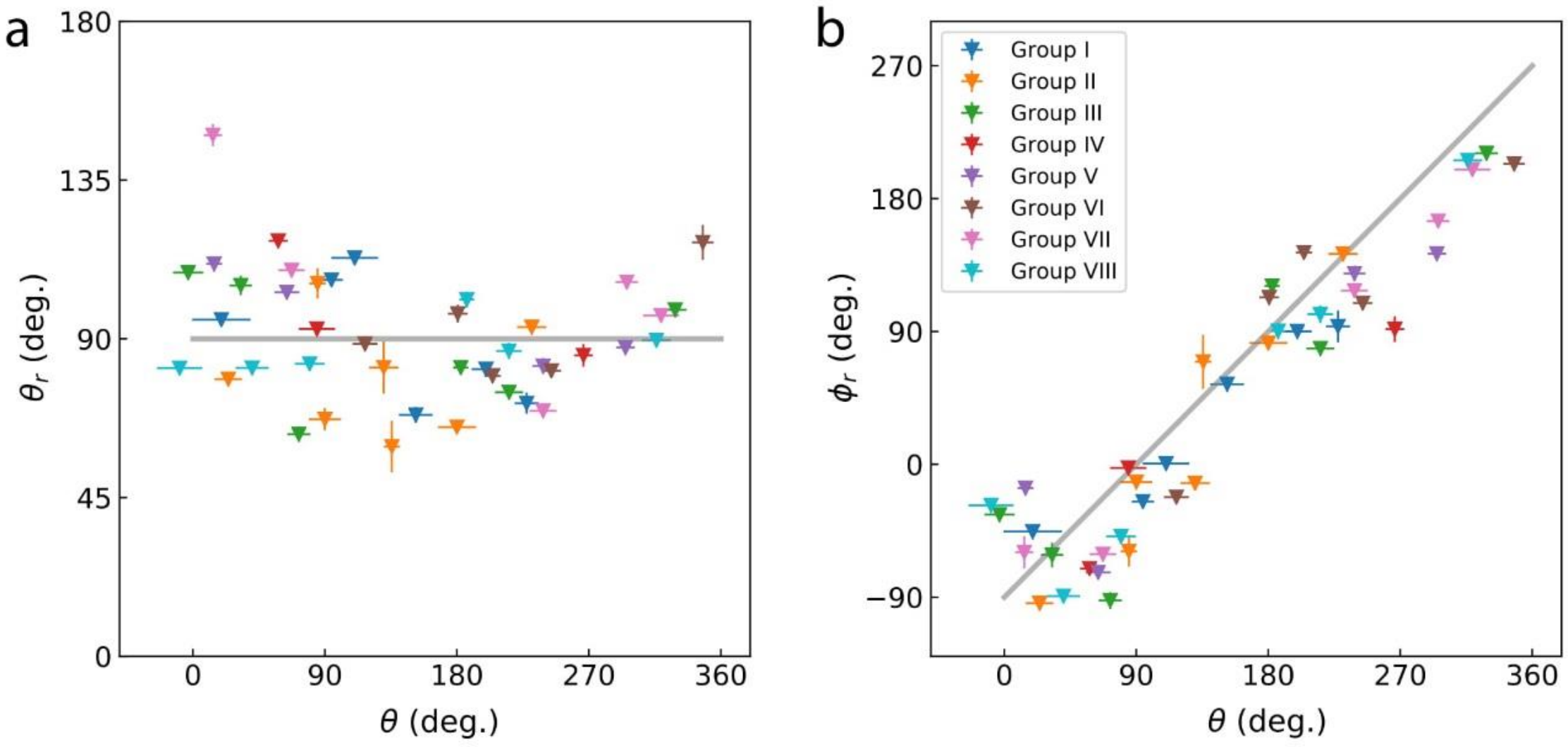

**Fig. S16 |** The relation of the rotation axes and angles. **a,** The polar angle $\theta_\mathrm{r}$ and **b,** the azimuthal angle $\phi_\mathrm{r}$ of NDs in the eight sets of experiments as functions of $\theta$. The grey line in **b** is the axisymmetric condition $\phi_\mathrm{r} = \theta - 90°$.

By assuming the axisymmetric condition, the gradient field can be written in polar coordinate as $(g_\rho, g_\theta) = (\tan\chi(\rho), 0)$. Then the deformed surface is reconstructed by integration, as $z(\rho) = \int_0^\rho \tan\chi(\rho')\, d\rho'$.

**Supplementary Information Note 7. Simulation of the gelatin deformation.**

Surface tension can play a major role in the mechanics of soft solids[12]. The dominance of surface tension or bulk elasticity can be evaluated using the elastocapillary length scale $L = \tau^0/E$, where $\tau^0$ is the surface tension and $E$ is the Young's modulus. When the scale of the deformation is much smaller than the elastocapillary length scale, surface tension dominates and flattens the surface. A linear-elastic model taking the effect of the surface tension into consideration was employed to simulate the deformation of the gelatin upon AFM tip indentation (schematic shown in Fig. S17a). In this model, the surface tension is introduced by adding a thin membrane of the same material ideally adhered to the bulk with negligibly thickness[13]. The system was simulated following methods given in Ref. 14. Young's modulus $E = 6.5$ kPa, Poisson's ratio $\nu = 0.5$ and surface tension $\tau^0 = 6.5\ \mathrm{mN\ m^{-1}}$ were employed as the simulation parameters. The possible shape of the AFM tip (cone with tip radius 150 nm, half-angle 8°) was assumed (from SEM). The simulated indentation force-depth profile agrees with the experiment results (Fig. S17b). The simulated result of force-depth profile using the Hertzian model[6,7] (taking Young's modulus $E = 36$ kPa and zero surface tension $\tau^0 = 0$) is also plotted in Fig. S17b (dashed line).

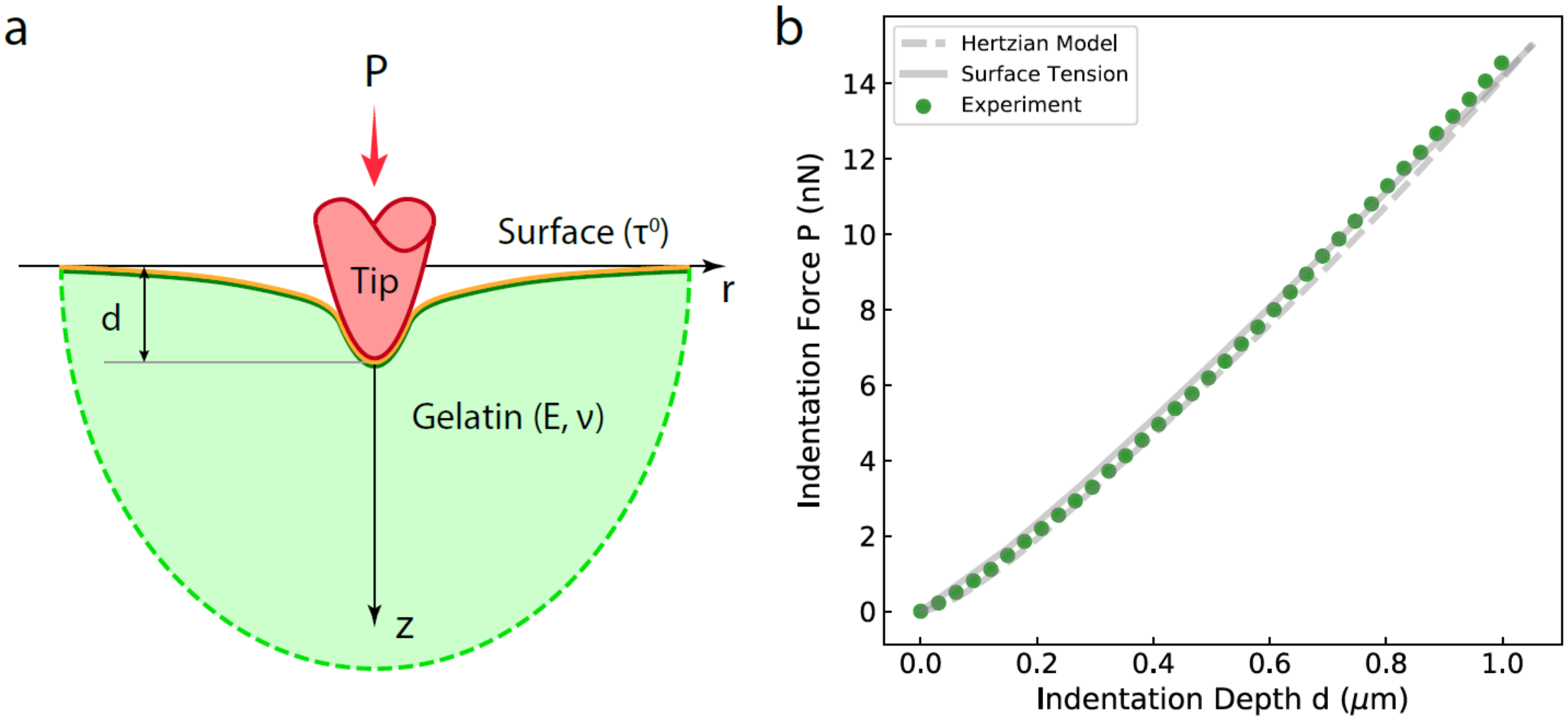

**Fig. S17 |** Simulation of the gelatin deformation. **a,** Schematic of the system with surface tension upon indentation. **b,** The force-depth profile of the indentation on the gelatin (green dots). The grey solid line and dashed line represent the simulation results with and without the surface tension, respectively.

## Supplementary References:

1. Doherty, M. W. *et al.* The nitrogen-vacancy colour centre in diamond. *Phys. Rep.* **528,** 1–45 (2013).
2. Rose, M. E. *Elementary theory of angular momentum*. (Courier Corporation, 1957).
3. Doherty, M. W. *et al.* Measuring the defect structure orientation of a single NV centre in diamond. *New J. Phys.* **16,** 063067 (2014).
4. Rondin, L. *et al.* Magnetometry with nitrogen-vacancy defects in diamond. *Reports Prog. Phys.* **77,** 056503 (2014).
5. Harker, M. & O'Leary, P. Regularized Reconstruction of a Surface from its Measured Gradient Field. *J. Math. Imaging Vis.* **51,** 46–70 (2015).
6. Hertz, H. Ueber die Berührung fester elastischer Körper. *J. für die reine und Angew. Math. (Crelle's Journal)* **1882,** 156–171 (1882).
7. Sneddon, I. N. The relation between load and penetration in the axisymmetric boussinesq problem for a punch of arbitrary profile. *Int. J. Eng. Sci.* **3,** 47–57 (1965).
8. Bowden, N., Huck, W. T. S., Paul, K. E. & Whitesides, G. M. The controlled formation of ordered, sinusoidal structures by plasma oxidation of an elastomeric polymer. *Appl. Phys. Lett.* **75,** 2557–2559 (1999).
9. Mills, K. L., Zhu, X., Takayama, S. & Thouless, M. D. The mechanical properties of a surface-modified layer on polydimethylsiloxane. *J. Mater. Res.* **23,** 37–48 (2008).
10. Nielson, G. M. A Method for Interpolating Scattered Data Based Upon a Minimum Norm Network. *Math. Comput.* **40,** 253 (1983).
11. Li, J. & Chou, T.-W. Elastic field of a thin-film/substrate system under an axisymmetric loading. *Int. J. Solids Struct.* **34,** 4463–4478 (1997).
12. Style, R. W., Jagota, A., Hui, C.-Y. & Dufresne, E. R. Elastocapillarity: Surface Tension and the Mechanics of Soft Solids. *Annu. Rev. Condens. Matter Phys.* **8,** 99–118 (2017).
13. Gurtin, M. E. & Ian Murdoch, A. A continuum theory of elastic material surfaces. *Arch. Ration. Mech. Anal.* **57,** 291–323 (1975).
14. Long, J. M. & Wang, G. F. Effects of surface tension on axisymmetric Hertzian contact problem. *Mech. Mater.* **56,** 65–70 (2013).